\documentclass[12pt,preprint]{aastex}
\usepackage{natbib,amssymb,amsmath,graphicx}
\shorttitle{zebra burst}
\shortauthors{Chen et al. }

\begin{document}

\title{Spatially and Spectrally Resolved Observations of a Zebra Pattern in
Solar Decimetric Radio Burst}

\author{Bin Chen$^{1,2}$, T. S. Bastian$^2$, D. E. Gary$^3$, and Ju Jing$^4$}
\affil{$^1$Department of Astronomy, University of Virginia, Charlottesville, VA 22904 USA\\
$^2$National Radio Astronomy Observatory, Charlottesville, VA 22903 USA\\
$^3$Center for Solar-Terrestrial Research, New Jersey Institute of Technology, Newark, NJ 07102 USA\\
$^4$Space Weather Research Laboratory, New Jersey Institute of Technology, Newark, NJ 07102 USA}

\begin{abstract}
We present the first interferometric observation of a zebra-pattern radio burst with simultaneous high spectral ($\approx 1$ MHz) and high time (20 ms) resolution. The Frequency-Agile Solar Radiotelescope (FASR) Subsystem Testbed (FST) and the Owens Valley Solar Array (OVSA) were used in parallel to observe the X1.5 flare on 14 December 2006. By using OVSA to calibrate the FST the source position of the zebra pattern can be located on the solar disk. With the help of multi-wavelength observations and a nonlinear force-free field (NLFFF) extrapolation, the zebra source is explored in relation to the magnetic field configuration. New constraints are placed on the source size and position as a function of frequency and time. We conclude that the zebra burst is consistent with a double-plasma resonance (DPR) model in which the radio emission occurs in resonance layers where the upper hybrid frequency is harmonically related to the electron cyclotron frequency in a coronal magnetic loop.
\end{abstract}

\section{INTRODUCTION}

Fine structures in the solar radio bursts - in both the time and frequency domains - have been studied for many years. They are believed to embody important information about charged particle acceleration processes, particle dynamics, and emission mechanisms \citep{1994SoPh..153..403F}. Many such fine structures - type III bursts and their variants, spike bursts, pulsations, fiber bursts - are believed to be the result of non-equilibrium processes in the coronal plasma. Zebra-pattern radio bursts (Slottje 1972; hereafter ``zebra pattern'' will be abbreviated to ``ZP'') are one of the most striking examples of such fine structures.

The observed properties of ZP radio bursts have been presented in detail in the review by \citet{2006SSRv..127..195C} and are reiterated briefly here. ZP bursts appear in radio dynamic spectra as closely-spaced, quasi-parallel bands of emission, typically ranging  from $\sim\!5-20$ in number but sometimes showing as many as 70.  They have been observed at meter wavelengths for decades \citep{1959Natur.184..887E,1972SoPh...25..210S,1975PhDT.........1K}; more recently, similar structures have been reported at decimeter and centimeter wavelengths. For the purposes of discussion we denote the instantaneous frequency of a single ZP emission band or stripe by $f_e$, the frequency bandwidth of a ZP emission band by $\Delta f_e$, the separation between adjacent ZP emission bands by $\Delta f_{s}$, the mean frequency of two adjacent emission bands as $f_m$, the mean frequency of the ZP emission bands as a whole by $\langle f_e\rangle$, and the overall frequency bandwidth occupied by ZP emission bands by $\Delta f_{tot}$. Generally,  $\Delta f_{tot}/\langle f_e\rangle$ decreases with frequency whereas $\Delta f_s/\langle f_e\rangle$ increases with frequency; the relative bandwidth of individual ZP emission bands $\Delta f_e/f_e$ shows no obvious trend with frequency and is typically $\lesssim\!1\%$ \citep{2006SSRv..127..195C}. There are few reports of the brightness temperature $T_B$ of ZP bursts. In those cases where such constraints are available, the brightness temperature is typically very high: \citet{1994SoPh..155..373C} estimated the $T_B$ of a metric ZP to be $\approx 10^{10}$ K with the source size constrained by the Nan\c cay Radioheliograph (NRH); In \citet{2003A&A...406.1071C}, a decimetric ZP that consisted of spiky superfine structures was estimated to have $T_B \gtrsim 10^{13}$ K by assuming the burst had the same source size as a spike burst; \citet{2005A&A...431.1037A} used the Siberian Solar Radio Telescope (SSRT) to observe a ZP burst at $\approx 5.7$ GHz, the highest frequency ever reported for ZP emission, which yielded a lower limit of $T_B\approx2\times10^7$ K,  the source size being $\lesssim 10''$. ZP bursts are typically observed during the impulsive and/or decay phases of the flares. They are typically polarized in the sense of the ordinary wave mode and the degree of polarization can be very high. The durations of ZP bursts can vary from a few minutes down to a few seconds at meter wavelengths to decimeter/centimeter wavelengths, respectively.  The narrow-band features, high degree of circular polarization, and indications of high brightness temperatures suggest that the corresponding emission mechanism is coherent.

ZP bursts often appear with the presence of type IV continuum emission (hereafter ``continuum''). Many other fine structures, including type III bursts, broad band pulsations (BBP), fiber bursts, and spikes accompany, or are associated with, ZP emission. There are also some rare examples where ZPs consist of pulsating superfine structures \citep{2007SoPh..241..127K, 2007SoPh..246..431C,2008SoPh..253..103K}, and they appear in fast-drift, type-III-like, absorption features  \citep{2009SoPh..255..273Z}, which could be related to fast electron beam injections into the magnetic trap.

There is no broadly accepted interpretation for ZP emission. Several types of models purport to explain the ZP phenomenon \citep[see][]{2006SSRv..127..195C,2009CEAB...33..281Z}. Most involve the growth and conversion of electrostatic wave modes to transverse modes. One type of model suggests that all the ZP stripes originate from the same discrete source, the dimensions of which are assumed to be small enough for the inhomogeneity of the plasma density and the magnetic field to be neglected. In these models, zebra stripes are assumed to be simultaneously generated at several harmonics of the local electron cyclotron frequency, due to nonlinear couplings of Bernstein waves with each other or with upper-hybrid waves (hereafter ``Bernstein models'') \citep[e.g.][]{1972SoPh...25..188R,1975SoPh...43..431Z,1975SoPh...44..461Z,1983SoPh...88..297Z}. 

A second type of ZP model is based on trapping upper hybrid Z mode waves in density inhomogeneities  \citep{2003ApJ...593.1195L}. The trap results in a discrete spectrum of eigenfrequencies. The model depends on the emission by many such discrete traps distributed over a larger volume. 

Models based on propagation phenomena have also been proposed. \citet{2006A&A...450..359B} and \citet{2006PlPhR..32..866L} suggest that coronal fine structure can behave as an optical filter or produce Bragg-like reflections, resulting in regular emission bands. Alternatively, interference between direct and reflected rays from a coherent source have been suggested \citep{2006SoPh..233..129L, 2010Ap&SS.325..251T}. 

Another class of models argues that ZPs are related to an extended source filled with energetic electrons. The different zebra stripes originate from different locations in the extended source, where resonance conditions are fulfilled.  The most popular model of this kind is the so-called double plasma resonance (DPR) model, first proposed by \citet{1975SoPh...43..431Z,1975SoPh...44..461Z}, and subsequently developed by several authors \citep[e.g.][]{1986ApJ...307..808W,2007SoPh..241..127K}. In this class of models, upper-hybrid waves are generated most efficiently at locations where the double plasma resonance occurs:
\begin{equation}
f_{UH}=(f^{2}_{pe}+f^{2}_{ce})^{1/2}=sf_{ce}
\end{equation}
where $f_{UH}$ is the upper-hybrid frequency, $f_{pe}$ is the electron plasma frequency, $f_{ce}$ is the electron cyclotron frequency, and $s$ is the harmonic number. The distribution of the DPR levels in the flare loop is determined by spatial gradients in the plasma density and magnetic field.

Finally, models are based on propagation of whistler wave packets \citep{1976SvA....20..582C,1990SoPh..130...75C}  across, or along, the magnetic trap where the energetic electrons generate Langmuir waves (hereafter ``whistler model''). ZPs are produced by coalescence of the Langmuir waves (\textit{l}) and whistlers (\textit{w}) through the process $l+w\longrightarrow t$, where \textit{t} stands for transverse waves that can be observed as emission near the local plasma frequency. They propose that ZPs as a whole are the manifestation of the ensemble of periodically generated whistler wave packets propagating in the magnetic trap, in which each zebra stripe corresponds to one propagating whistler wave packet. In this way, zebra stripes can be separated regularly from each other in height (and emit at different frequencies) by a distance determined by the whistler propagation velocity and time interval of generating the whistlers.

Both \citet{2006SSRv..127..195C} and \citet{2009CEAB...33..281Z} discuss each of these models and summarize their strengths and weaknesses. Many display significant theoretical shortcomings. In light of these shortcomings,  Chernov favors whistler models. In contrast, Zlotnik favors DPR models. We will therefore direct most of our attention toward the last two classes of models - DPR and whistler models - in subsequent discussion. 

As for the type IV continuum emission associated with ZP bursts, it is assumed by most models that it arises from fast electrons trapped magnetically in the coronal loops. However, the relationship between the continuum and ZP emission varies from model to model. In the Bernstein and DPR models, it is suggested that the only difference between their formation is whether these trapped electrons favor conditions for creating the zebra pattern or not, which is most likely the forms of anisotropic momentum distributions. In the whistler model, the ZP is formed through interactions of the plasma waves that can result in the type IV continuum. The propagation models, however, suggest that either the ZP sources reside in the continuum source \citep{2006SoPh..233..129L}, or ZP emission as well as the ``background'' continuum as a whole are just  type IV continuum emission modulated by the inhomogeneous medium \citep{2006A&A...450..359B, 2006PlPhR..32..866L}. However, because of the lack of spatial information, the physical relationship between the emission sources of ZPs and type IV continuum is still not known.

The means of placing meaningful constraints on models for ZP and continuum emission has been limited by the unavailability of spatial resolution at each of the frequencies and times recorded by the dynamic spectrum. Interferometric observations of ZP emission at even a single frequency are relatively few in number. Several examples are provided by fixed-frequency observations of ZP by the NRH in combination with observations made by a spectrometer \citep[e.g.][]{1994SoPh..155..373C, 1998A&A...334..314C, 2001RaSc...36.1745C, 2003A&A...410.1001A, 2005PlPhR..31..314C}. Other examples have been provided by \citet{2005A&A...431.1037A} and \citet{2005A&A...437.1047C, 2006SoPh..237..397C} who combined interferometric observations of ZP made by the SSRT at 5.7 GHz with spectroscopic observations obtained by the Chinese Solar Broadband Radio Spectrometer (SBRS/Huairou).

In the present paper we describe the first interferometric observations of ZP emission for which both high-time resolution and high-spectral resolution observations are simultaneously available over a significant frequency bandwidth. The relevant instrumentation is described in \S2. The ZP event and its analysis are described in \S3. The event is placed in a physical context in \S4, where we argue that the data are consistent with a DPR model. We conclude in \S5.

\section{Instrumentation}
\label{sect:instr}
The observations were obtained by the Frequency-Agile Solar Radiotelescope (FASR) Subsystem Testbed (FST) \citep{2007PASP..119..303L}. FASR is a next generation solar radio telescope \citep{2004ASSL..314.....G} designed to provide simultaneous imaging and spectroscopic observations over a large bandwidth, with high angular, time, and spectral resolutions. The FST is a prototype and testbed system for FASR. As such, it is the first system with the ability to combine Nyquist-limited high time and frequency resolution with interferometric ability to locate sources \citep{2007PASP..119..303L}.

\begin{figure}
\begin{center}
\includegraphics[width=0.8\textwidth]{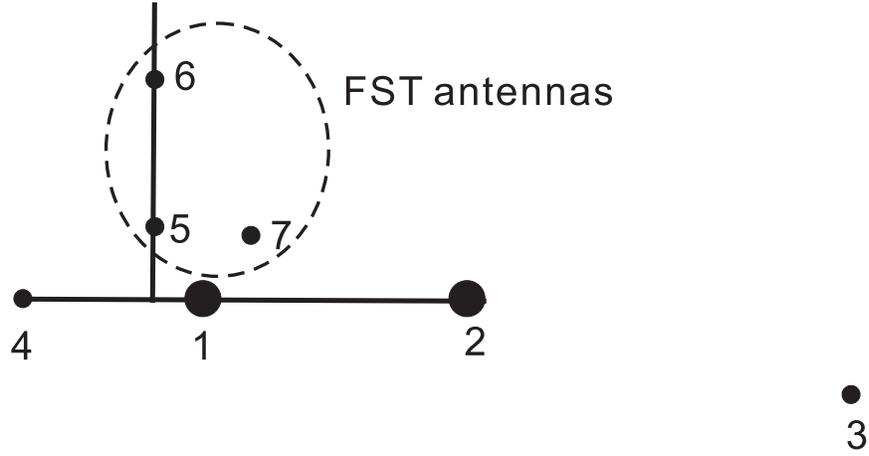}
\caption{The nearly right triangle antenna configuration of FASR Subsystem Testbed, which consists of three 1.8-m antennas numbered 5, 6, and 7 of the Owens Valley Solar Array \citep[adapted from][]{2007PhDT....Liu}.}\label{fig:ants}
\end{center}
\end{figure}

The FST uses the existing antenna system of the Owens Valley Solar Array (OVSA). OVSA \citep{1994ApJ...420..903G} is a solar-dedicated interferometric array that is composed of two 27-meter antennas and five 1.8-meter antennas. OVSA can observe the Sun at up to 86 frequencies in the range 1-18 GHz. The FST employs three of the 1.8-meter OVSA antennas (antenna number 5, 6 and 7) as shown in Figure \ref{fig:ants}. The longest baseline, between antennas 6 and 7, is nearly 280 meters which yields a minimum fringe spacing of 221$''$ at 1 GHz. \citet{2007PASP..119..303L} describe the FST system configuration in detail, and we briefly reiterate it here. A radio frequency (RF) splitter divides the output signals of the three OVSA/FST antennas to simultaneously feed the OVSA receivers and the FST. Thus, the FST can be used in parallel with OVSA, observing the same source as OVSA,  without affecting OVSA's normal operation. FST signals are amplified and transmitted by a broadband optical fiber to a block down-converter. The 1-5 GHz or 5-9 GHz band of RF signal from each antenna is selected and down-converted to 1-5 GHz as necessary. The spectral-line down-converter is used to select a single-sideband, 500 MHz band which is tunable anywhere within the 1-5 GHz band. The selected 500 MHz band is further down-converted it to an intermediate frequency (IF) band of 500-1000 MHz. 
The 500 MHz IF band is sampled at 1 Gsps and the data are written to disk. The full-resolution time-domain data are then correlated offline using a software correlator (written in IDL) to produce amplitude and phase spectra on the three interferometric baselines. For daily solar observing, the system employs a time resolution of $\approx\!20$ ms and a frequency resolution of $0.98$ MHz.

\begin{figure}
\begin{center}
\includegraphics[width=0.85\textwidth]{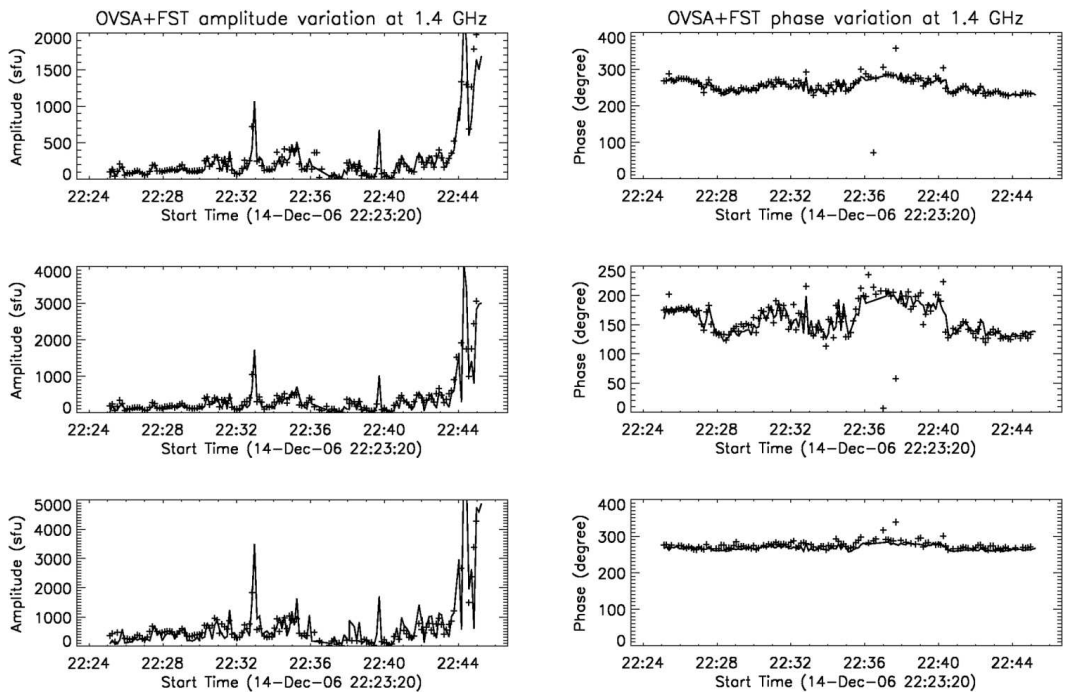}
\caption{Comparison of OVSA and FST amplitudes and phases (OVSA: plus; FST: solid line). From top to bottom: baselines of antenna 5-6, 6-7, 7-5, respectively.}\label{fig:comp}
\end{center}
\end{figure}

The small size of the FST antennas precludes calibration of the FST against sidereal radio sources. However, since the event was observed by both FST and OVSA in parallel, we can make use of the OVSA observations to cross-calibrate FST. OVSA is calibrated against sidereal standards in order to determine the complex gain of each antenna. However, OVSA only samples the RF spectrum at 1.2 and 1.4 GHz. Moreover, the 1.2 GHz OVSA data were found to be corrupted and were therefore unusable. Therefore, the FST 1.4 GHz data were averaged in frequency and time to match the OVSA data. The calibrated OVSA amplitudes and phases of antenna baselines 5-6, 6-7, and 5-7 were then compared to those measured by FST, and the FST amplitudes and phases were corrected accordingly. As Figure \ref{fig:comp} shows, they agree with each other quite well after the cross-calibration. However, this process doesn't directly allow bandpass calibration of FST. Based on an examination of  broadband continuum emissions that occupy the entire frequency band, we conclude that the FST has approximately linear bandpass patterns in phase with good stability in time on all three baselines. We therefore applied linear fits as the bandpass correction to the phases. 

\section{Observations}

\begin{figure}
\begin{center}
\includegraphics[width=0.75\textwidth]{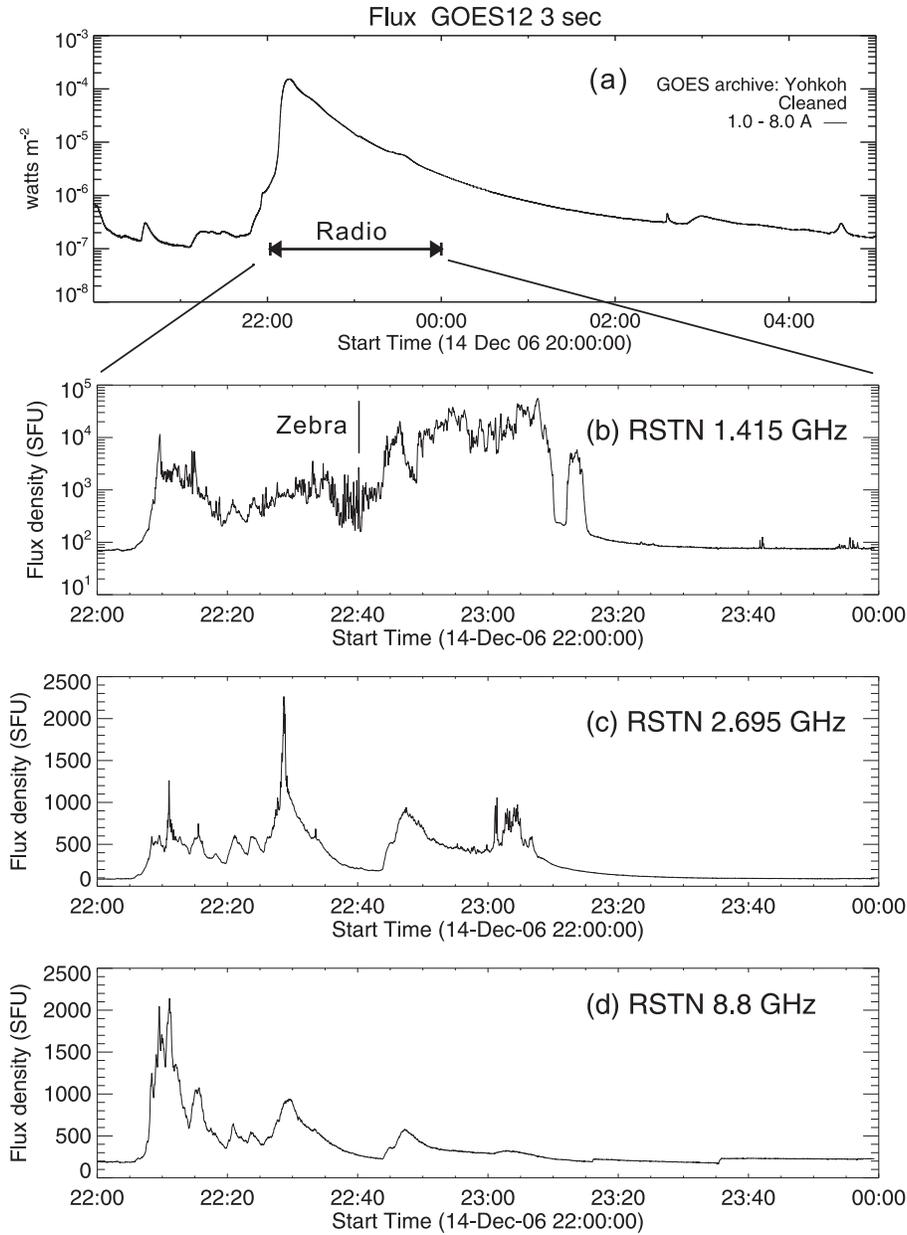}
\end{center}
\caption{GOES and RSTN time profiles of the 2006 December 14 flare. Note the difference in scale between the intense 1.415 GHz emission and that at 2.695 and 8.8 GHz. The time of the ZP event is marked by the vertical line in (b), at about 22:40 UT, during the decay phase.}\label{fig:goes_rstn}
\end{figure}

The FST observed the powerful GOES class X1.5 flare that occurred on 2006 Dec 14 in NOAA/USAF active region 10930 at S06W46, the site of a X3.4 flare the previous day \citep[see, e.g.,][]{2007PASJ...59S.785S}. The flare on Dec 14 was accompanied by a fast halo CME and a solar energetic particle event. Figure \ref{fig:goes_rstn}a shows the GOES SXR light curve; the flare started at 21:07 UT, peaked at 22:15 UT, and ended at around 04:00 UT on December 15. The radio time profiles at 1.415 GHz, 2.695 GHz, and 8.8 GHz obtained by the Radio Solar Telescope Network (RSTN, operated by the U.S. Air Force) are shown in Figure \ref{fig:goes_rstn}b$-$d. Note the difference in scale between the intense 1.415 GHz emission and that at 2.695 and 8.8 GHz. 

\begin{figure}
\begin{center}
\includegraphics[width=0.85\textwidth]{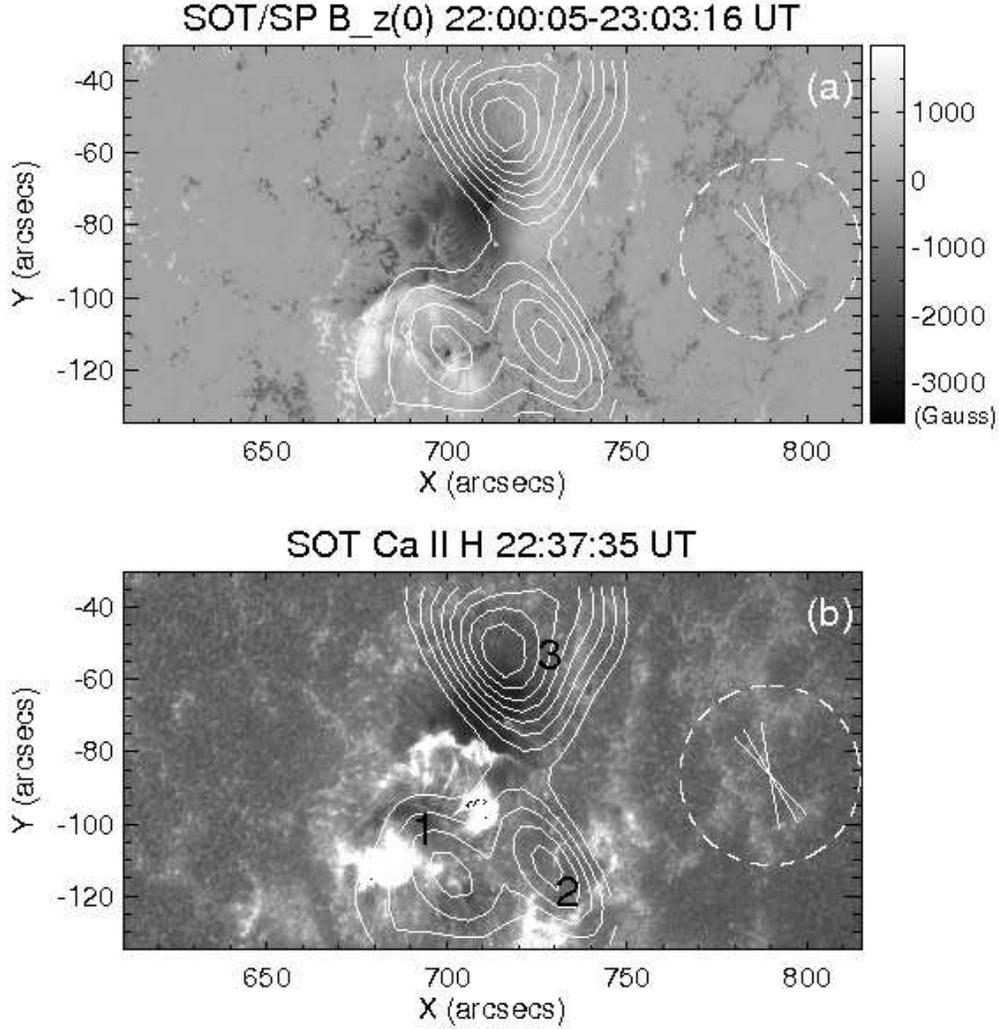}
\caption{Longitudinal photometric magnetogram $B_z(0)$ (a) and an example of a Ca II H image at 22:37:35 UT (b) by SOT. North is up and west is to the right. The magnetogram was observed from 22:00:05 to 23:03:16 for about one hour. The intersection of the three interferometric fringes of FST denote the location of the ZP emission centroid (observed around 22:40 UT), with a dashed circle representing the $\approx 50''$ apparent source size obtained in \S \ref{obs:srcsize}. The error of the source location is between 1.3$''$ and 3.5$''$ depending on direction. The contours are the OVSA 4.6$-$6.2 GHz map at the same time of the ZP (the levels have an increment of 10\% of the maximum). The numbers ``1'', ``2'', and ``3'' in (b) mark the locations of the three major Ca II H bright regions seen in Figure \ref{fig:ca2h_n_xrt}a-c.}\label{fig:SOTSP}
\end{center}
\end{figure}

The two X-class flares on 2006 Dec 13 and 2006 Dec 14 were observed by the X-ray Telescope (XRT) \citep{2007SoPh..243...63G} and the Solar Optical Telescope (SOT) \citep{2008SoPh..249..167T} aboard {\sl Hinode} \citep{2007SoPh..243....3K}. Most of the XRT images were taken with a $512''\times 512''$ field of view (FOV) and a cadence of 60 s. Vector photospheric magnetograms were obtained by the SOT SpectroPolarimeter (SP). The slit length and width of SOT/SP are $164''$ and $0.16''$, respectively. For each magnetogram with a $297''\times 164''$ FOV, the slit scanned from east to west for about one hour, with a resolution of $0.297''$ and $0.320''$ in the east-west and north-south directions, respectively. The SOT Ca II H band samples the chromospheric structure at a very high spatial resolution ($0.1”$ per pixel). Like H$\alpha$ this band is sensitive to plasma heating by precipitating electrons and/or thermal conduction. The images were taken with a $218''\times 109''$ FOV with a cadence of 120 s. Figure \ref{fig:SOTSP}a and b give the longitudinal magnetogram at the photospheric level $B_z(0)$ observed from 22:00:05 to 23:03:16 UT, and an example of a Ca II H image at 22:37:35 UT.

\begin{figure}
\begin{center}
\includegraphics[width=0.85\textwidth]{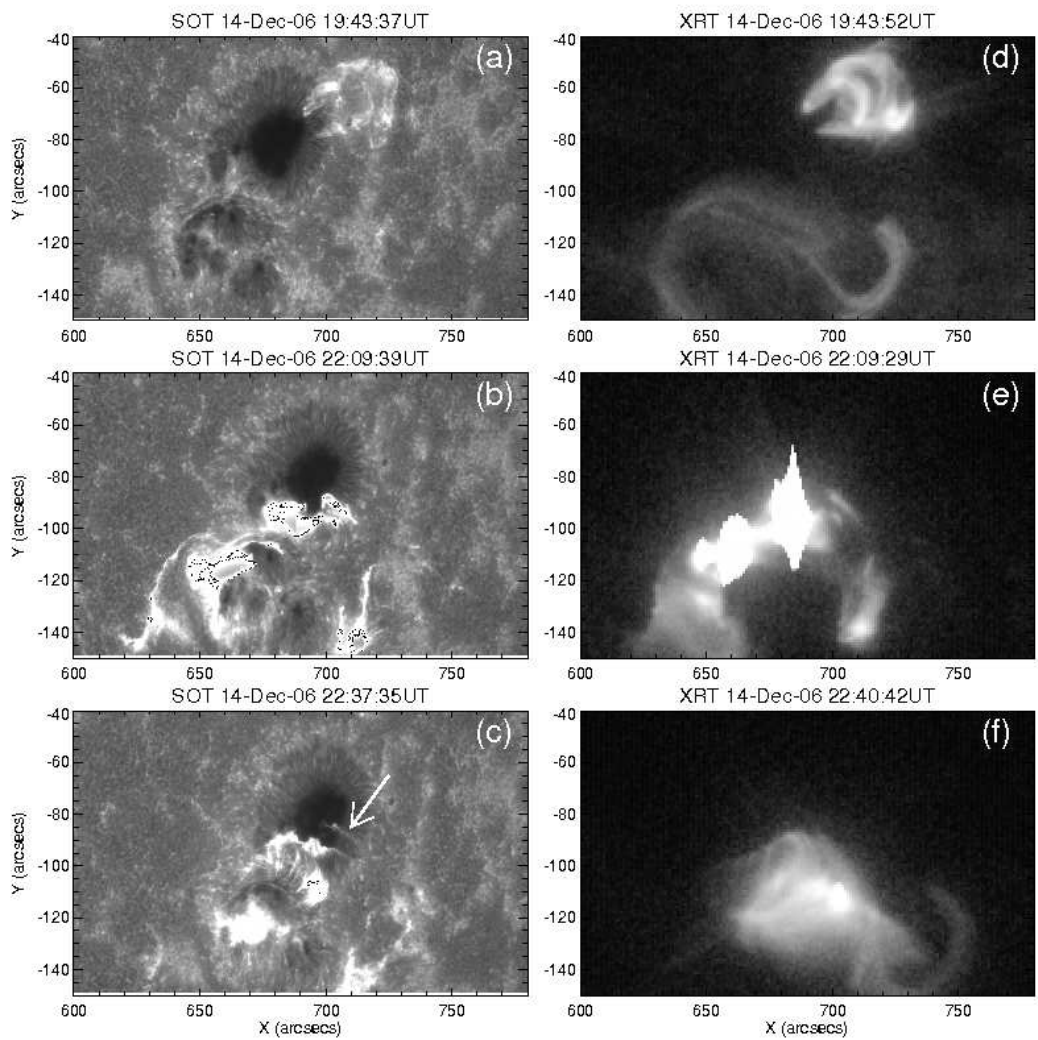}
\caption{Hinode SOT Ca II H and XRT observations of the 2004 December 14 flare showing the chromospheric and coronal evolution. The left column: Hinode Ca II H images at 19:43:37, 22:09:39, and 22:37:35 UT. The right column: Hinode XRT images at 19:43:52, 22:09:29, and 22:40:42 UT. They correspond to the times prior to the flare, during the flare maximum, and of the occurrence of ZP in the post-flare phase. The white arrow in (c) shows the subtle Ca II H brightening that may be magnetically associated with the ZP source. }\label{fig:ca2h_n_xrt}
\end{center}
\end{figure}

Figure \ref{fig:ca2h_n_xrt} shows the Ca II H (Figure \ref{fig:ca2h_n_xrt}a$-$c) and XRT (Figure \ref{fig:ca2h_n_xrt}d$-$f) images at times prior to the flare (a and d), during the flare maximum (b and e), and during the decay phase (c and f) at the time of the ZP. The core flare region (marked as region ``1'' in Figure \ref{fig:SOTSP}b) is located between the two sunspots with opposite polarity. According to \citet{2007PASJ...59S.785S}, the flare loops inside this region evolve from highly sheared in the pre-flare phase into less sheared in the post-flare phase, based on the XRT observations. Except for the flare core region, there are two other major bright regions in the Ca II H images, where the brightness of soft X-ray loops is enhanced at the same time, as shown in the XRT images. One is located about $50''$ to the west of the core region (region ``2'' in Figure \ref{fig:SOTSP}b), and another is located to the north-west near the larger sunspot with negative polarity (region ``3'' in Figure \ref{fig:SOTSP}b). 

A more subtle Ca II H brightening is observed at the time of the ZP (at about 22:40 UT), indicated by a white arrow in Figure \ref{fig:ca2h_n_xrt}c). This brightening appeared in Ca II H images at around 22:10 UT (near the flare maximum) and persisted for more than an hour. 

\subsection{FST ZP Observations}
\label{obs:fst}

The FST observed the 2006 Dec 14 flare over a frequency range of 1.0$-$1.5 GHz. The instrument observed both right- and left-circularly polarized radiation, switching between the two polarizations every 4 s. A spectrum of 511 channels across the 500 MHz bandwidth was produced every $\approx\!20$ ms. The 2006 Dec 14 flare produced a high level of type IV burst activity in the post-flare phase for more than one hour. A rich variety of fine structures was observed in the 1.0$-$1.5 GHz band, including ZP, fiber bursts, pulsations, and others. The time of the ZP event is marked by the vertical line in Figure \ref{fig:goes_rstn}b, at about 22:40 UT, during the decay phase.

\subsubsection{Total Power Dynamic Spectrum}

\begin{figure}
\begin{center}
\includegraphics[width=0.85\textwidth]{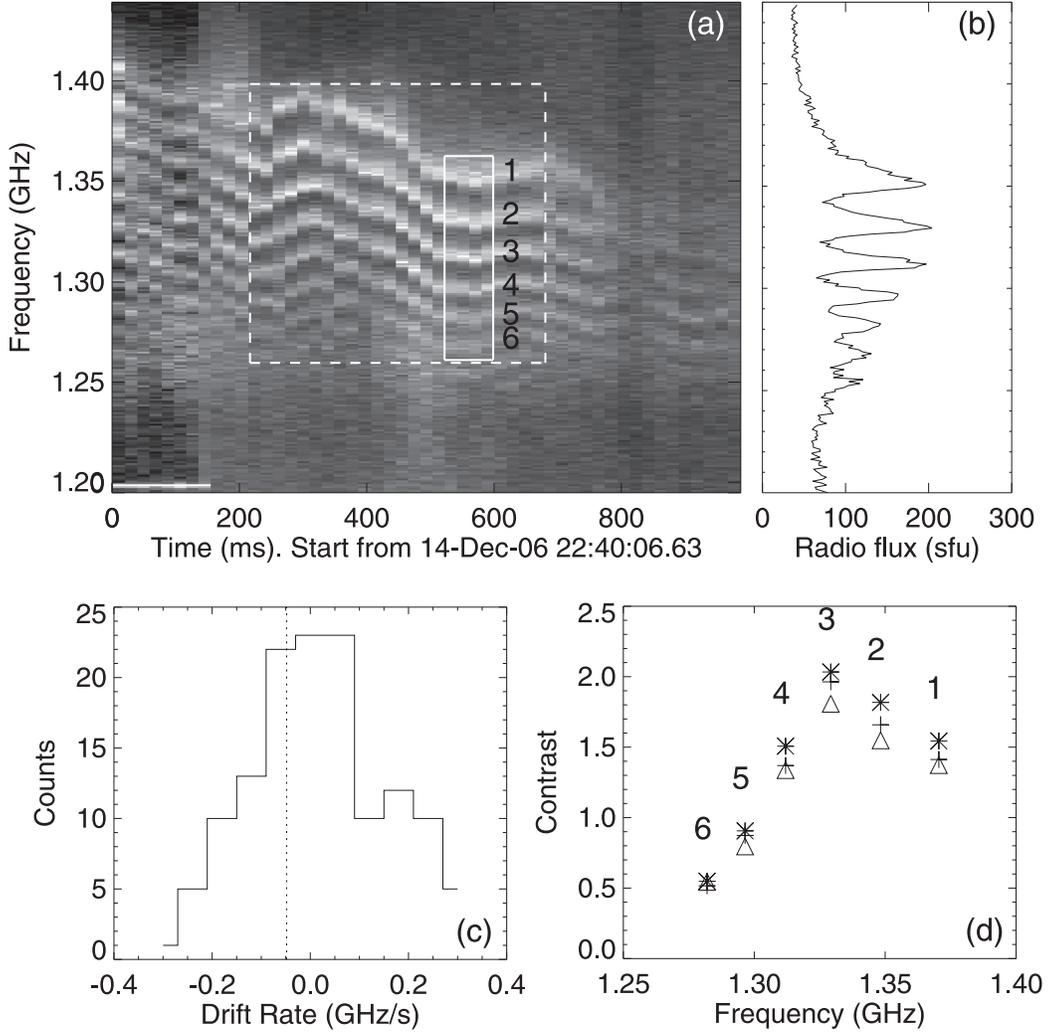} \\
\caption{(a): A zebra-pattern structure observed at around 22:40 UT on 2006 December 14. Six successive strong stripes with decreasing frequencies are marked by numbers 1 to 6. (b): Frequency profile of zebra-pattern structure, averaged in the time denoted by the small solid box (between 22:40:07.18 UT and 22:40:07.26 UT). (c): Histogram of drift rates of zebra stripes in the large dashed box. Vertical line indicates the mean drift rate is about -50 MHz/s. (d): Intensity contrasts $P=(I_{on}-I_{off})/I_{off}$ at the six zebra stripes. Pluses, stars, and triangles denote respective contrasts of baseline 5-6, 6-7, and 7-5. }\label{fig:zebra}
\end{center}
\end{figure}

Figure \ref{fig:zebra}a shows a dynamic spectrum of roughly 1 s of total power data near 22:40:07 UT. A striking ZP radio burst is present.  The data are right-circularly polarized (RCP); the left-circularly polarized (LCP) observations made just prior to those presented here showed no ZP emission. We conclude that the ZP burst is highly right-circularly polarized. A precise measure of the degree of polarization is not possible, however, because the noise in the LCP observations is dominated by polarization leakage from the RCP channel. The ZP burst shows as many as 12 distinct bands or stripes superposed on a type IV-like continuum background centered on $\langle f_e \rangle \approx 1.32$ GHz and with a frequency bandwidth $\Delta f_{tot}$ of up to 150 MHz, or $\Delta f_{tot}/\langle f_e\rangle \gtrsim 10\%$.  All of the zebra stripes drift in frequency together irregularly with time. Figure \ref{fig:zebra}c shows a histogram of the drift rates of the zebra stripes in the dashed box marked in Figure \ref{fig:zebra}a. An average drift rate of about $-$50 MHz~s$^{-1}$ is indicated by the vertical line. The overall trend is to drift from higher to lower frequencies. The drift rates of the zebra stripes are mainly between $-100$ MHz~s$^{-1}$ to 100 MHz~s$^{-1}$, comparable to, but generally slower than, those of the so-called ``fiber burst'' in the same frequency range \citep{1982ITABO..53.....E}.

The frequency profile of the ZP emission (averaged in the time denoted by the small solid box in Figure \ref{fig:zebra}a) is shown in Figure \ref{fig:zebra}b. The contrast of each stripe is defined by $(I_{on}-I_{off})/I_{off}$ (where $I_{on}$ and $I_{off}$ are the intensities ``on'' and ``off'' a zebra stripe, respectively) and is shown in Figure \ref{fig:zebra}d. The relative frequency separations between adjacent zebra stripes $\Delta f_s/f_m$ in the dashed box of Figure \ref{fig:zebra}a are shown in Figure \ref{fig:zb_spacing}a as diamonds. They are color-coded in time from blue to red. One can clearly see that $\Delta f_s/f_m$ increases with frequency. This phenomenon is commonly seen in zebra patterns reported by other authors in both the meter and decimeter wavelength range \citep[e.g.][]{2005PlPhR..31..314C,2005A&A...437.1047C}.

\subsubsection{Apparent ZP Source Size}
\label{obs:srcsize}
With only three antenna baselines it is not possible to image the ZP source. However, we are able to constrain the source size using the visibility amplitudes as a function of antenna baseline. We assume that we can characterize the source brightness distribution as a symmetrical Gaussian. The visibility function is then likewise a Gaussian and a simple model, characterized by a single spatial scale, can be fit to the normalized visibility amplitudes as a function of spatial frequency. We have fit the source at frequencies ``on'' the bright zebra stripes (hereafter ``on-stripe source'') and at frequencies ``off'' the zebra stripes; that is, in between the zebra stripes (hereafter ``off-stripe source''). As is shown in Figure \ref{fig:srcsize}a, a simple Gaussian model adequately fits both the on- and off-stripe sources. The visibility amplitudes of the three baselines have been normalized by the total power measured by each antenna. Figure \ref{fig:srcsize}b shows the corresponding Gaussian FWHMs of the on- and off-stripe sources in the spatial domain for the six zebra stripes in Figure \ref{fig:zebra}a. They both have values near $50''$, but the sizes of the off-stripe sources are systematically larger than those of the on-stripe sources by $\approx\!10''$, except for the stripes at lower frequencies where the contrast is low (see Figure \ref{fig:zebra}d), a point to which we return in \S3.1.3. 

The uncertainty of the source size estimation depends on the accuracy of normalized amplitude measurements. Within the context of the model a difference of 20\% in the inferred FWHM of the on- and off-stripe sources implies that the normalized amplitude of the off-stripe source on the longest antenna baseline (antennas 6-7) is lower than that of the on-stripe source by $\gtrsim 10\%$. The normalized amplitude measurements used in our fitting are averaged values along each zebra stripe. Therefore, the systematic errors can be represented by the statistical standard deviation of the mean of the sample, which are respectively $\lesssim~1\%$ and $\lesssim~2\%$ for the on- and off-stripe sources. We therefore regard errors as large as 10\%  as unlikely and conclude that the on-stripe source is marginally more compact than the off-stripe source. Qualitatively similar results were obtained by \citet{1994SoPh..155..373C} using one-dimensional NRH observations at a much lower frequency of 164 MHz.

Note that both sources are likely strongly affected by scattering.  \citet{1994ApJ...426..774B} showed that scattering by the overlying inhomogeneous corona can play an important role in modifying the angular structure of the emission source at wavelengths longer than a few centimeters. The estimated source sizes of $\approx\!50''$ are consistent with angular scattering and the intrinsic source sizes could be significantly smaller. Given an apparent source size of $50''$, the lower limit of the brightness temperature of ZP source can be estimated to be $\approx\!10^9$ K.

\subsubsection{Relative Locations of the ZP and Continuum Sources}

We now ask whether the on- and off-stripe emissions originate from two different source locations. If they are indeed separated from each other spatially, how are they related to each other? 
Second, how do the source locations vary with time?
Finally, do the on- and off-stripe source locations have any significant dependence on frequency? In other words, does the radiation at different frequencies come from different spatial locations or not?

In interferometry, the interferometric phase is a direct response to the spatial information of the radiation source. We first consider the ``dynamic phase spectrum'' of the ZP emission for the three baselines to gain a qualitative impression (Figure \ref{fig:zb_phz}). For baselines 5-6 and 6-7, zebra stripes can be distinguished as darker colors compared with the background continuum. This means that the on- and off-stripe phases are measurably different. In addition, the phase difference seems larger in the upper-left region (higher frequencies and earlier times) than later in the event. For baseline 7-5, however, no phase differences are evident between on-stripe and off-stripe emissions.

In order to characterize on- and off-stripe phases quantitatively, we employ ``phasor diagrams''. In a phasor diagram, the amplitude and phase of each measurement are displayed in a polar plot. We have made phase measurements for both the on- and off-stripe emissions of the spectral fragment indicated by the dashed box in Figure \ref{fig:zebra}a, which includes six zebra stripes in an interval of 0.48 s with 24 consecutive observations (from 22:40:06.86 UT to 22:40:07.34 UT). The rms phase noise of a single data point (duration 20 ms and bandwidth roughly 1 MHz), is about 5 degrees, too large for tracking the phase variations in time and frequency among individual measurements. To increase the signal-to-noise ratio we averaged the data as follows. First, we obtained the on- and off-stripe phases by averaging the three frequencies about the flux maximum/minimum at each time. Second, we considered averages of the phases ``along''  the on- and off-stripe positions (in time) and ``across''  the six on- and off-stripe positions (in frequency). By averaging along the zebra stripes, we constrain the phase variations of the on- and off-stripe sources as a function of frequency, so that the spatial distribution of the ZP source over frequency can be revealed. Implicit in this treatment is the assumption that the ZP source is likely to maintain the same structure as a function of frequency during the brief averaging time. Similarly, by averaging the on- and off-stripe phases in frequency, we track the phase variation of the ZP source with time, so that the evolution of the source centroid location with time can be seen, considering that different stripes of the ZP show the same drift motion according to the dynamic spectra. This approach allowed us to reduce the statistical phase error by a factor of several, and to show systematic variations in the on- and off-stripe phases in the phasor diagrams with increased signal-to-noise ratio.

Figure \ref{fig:phasor_t} shows the results for the on-stripe (pluses) and off-stripe (triangles) emission after averaging across the six zebra stripes. The pluses and triangles are colored from dark blue to red to indicate the variation of amplitudes and phases in time. The dashed lines in each panel represent increments of $5^\circ$ in phase. It can be seen that the on-stripe phases of baseline 5-6 and 6-7 drift by $\approx\!9^\circ$ within 0.48 s towards the off-stripe phases, while the off-stripe phases show no evident drift. No significant phase drift is seen for baseline 7-5, thus the direction of the spatial drift is nearly along the 7-5 interferometric fringe, which is NE-SW at the time of observation. By applying the fringe spacings (which are respectively 274$''$, 176$''$, and 439$''$ for baselines 5-6, 6-7 and 7-5 at the time of observation) and orientations of the three FST baselines (see Figure \ref{fig:SOTSP}b), this amount of phase drift in time can be translated into a spatial drift of $15.6\pm6.5''$ from NE to SW on the solar disk, corresponding to a projected drift velocity of $2.5\pm1.0\times10^9$ cm$~s^{-1}$ ($\approx\!0.1c$). The estimated spatial error is based on the rms error of the FST phase measurements ($\approx\!5^\circ$), the total number of data points averaged, and triangulation of the three fringes. The fringe orientations of the three baselines yield the error to be orientationally dependent, which is larger in the NE-SW direction than the SE-NW direction.

In addition, there is an evident difference in mean phase of about $6^\circ$ between the on- and off-stripe emissions for baseline 5-6 and 6-7, shown by the arrows in Figure \ref{fig:phasor_t}. That means the on- and off-stripe emission are separated in space by an average of $8\pm0.9''$ in the NE-SW direction. 

In the two bottom panels of Figure \ref{fig:phasor_t}  the ZP phases shown in the top-left and top-middle panels (baselines 5-6 and 6-7) are plotted as a function of time. The solid line, repeated in each panel, represents the ZP frequency as a function of time (note that the frequency scale on the right axis is reversed  -  with frequency decreasing from high to low values). Although the variations in time are irregular, there is a rather good correlation between the phase and the frequency; stated another way, there is a good correlation between the ZP source centroid position and the mean ZP frequency. 

Figure \ref{fig:phasor_f} shows similar phasor diagrams. Here, however, the data are averaged along the zebra stripes, instead of across the zebra stripes as above, and the amplitude and phase variations are shown as a function of frequency. The data are color-coded from black to red for stripes with decreasing frequencies. The pluses and triangles denote the on- and off-stripes sources, respectively. The amplitudes of the off-stripe source barely change, but those of the on-stripe source drop with decreasing frequency. There is still little change in phase for baseline 7-5, but both the on- and off-stripe phases shift by several degrees monotonically across the six stripes. The shifts of on-stripe phases seem to be slightly larger than those of the off-stripe phases. Therefore, it appears that the source centroid positions of both the on- and off-stripe sources show a systematic displacement with frequency, which is respectively $14.6\pm2.8''$ and $8.3\pm2.8''$ across the six zebra stripes from NE to SW.

The spatial difference between the on-stripe and off-stripe emissions demonstrates that there are indeed two spatially separated sources that contribute to the event, namely, the zebra-stripe emission and the background continuum source. This, too, has been noted in previous examples of ZP events using one-dimensional NRH observations at much lower frequencies \citep[and references therein]{2006SSRv..127..195C}. Care must be taken in further interpretation of the phasor diagrams. The off-stripe emission is dominated by the continuum source. However, the on-stripe emission has contributions from both the continuum and ZP sources. Therefore, a change of relative intensities of the zebra and continuum sources can result in an apparent position shift of the emission centroid measured at on-stripe frequencies. The phase drift of the on-stripe emission in time (Figure \ref{fig:phasor_t}) is not in question in this respect because the amplitudes barely change in time along a given stripe. However, there is an obvious change of on-stripe amplitude with frequency at a given time - the amplitude decreases from high to low frequencies and the phase tends toward the value of the off-stripe emission (Figure \ref{fig:phasor_f}), which could be caused by the intensity modulation itself. In order to correct for this effect, we vector-subtract the complex amplitudes of the off-stripe source from those of the on-stripe source in the phasor diagram, and plot the actual phases of the zebra source as ``X'' symbols in the first and second panels of the bottom row in Figure \ref{fig:phasor_f}. After the correction, the ZP phases of the baseline 5-6 and 6-7 still show monotonic displacements as a function of frequency, but now they become comparable to those of the continuum (off-stripe) phases.
Again converting to angular displacements, the corrected ZP source centroid has a displacement of $8.5\pm2.8''$ across the six zebra stripes with a direction from NE to SW, which is smaller than the value we obtained previously for the on-stripe source ($14.6\pm2.8''$) but comparable to that of the continuum (off-stripe) source ($8.3\pm2.8''$).

\begin{figure}
\begin{center}
\includegraphics[width=12cm,keepaspectratio]{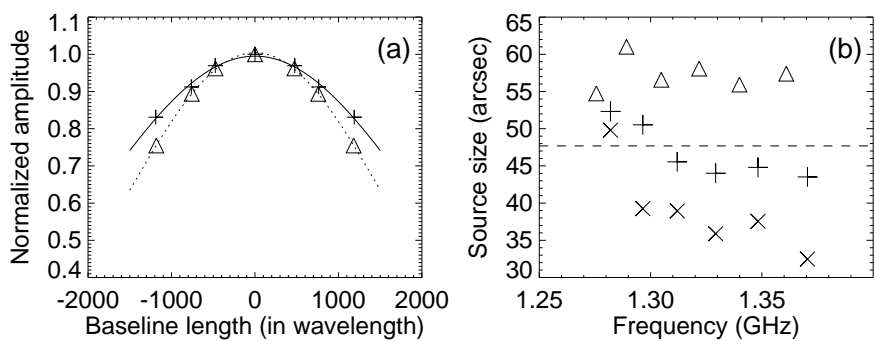}
\caption{(a): An example of Gaussian fits of the relative visibility amplitudes on one zebra stripe. Pluses and triangles denote the on- and off-stripe sources respectively. The seven data points for each fit are the relative visibility amplitudes for the three FST baselines (both positive and negative, plus one total power/zero-spacing amplitude). A narrower Gaussian in this visibility plot implies a spatially larger source. (b): Source size estimations on the six zebra stripes in Figure \ref{fig:zebra}a. The ``X'' symbols denote the zebra source sizes after removing the contribution from the continuum (off-stripe) source. The average value of the on- and off-stripe source sizes is given by the horizontal dashed line. }\label{fig:srcsize}
\end{center}
\end{figure}

\begin{figure}
\begin{center}
\includegraphics[width=0.85\textwidth]{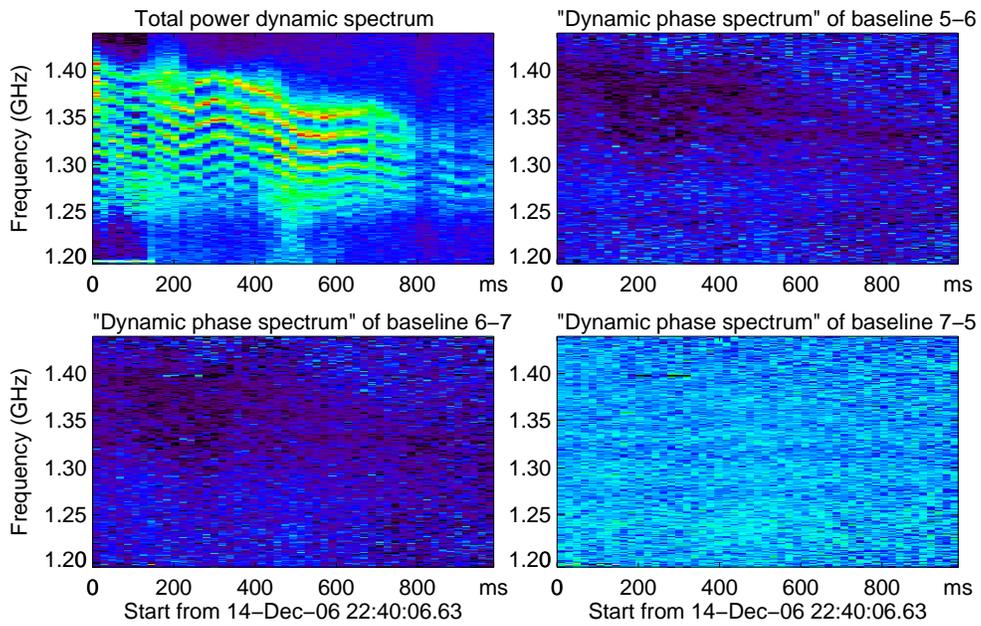}
\caption{The total power dynamic spectrum as well as the ``dynamic phase spectrum'' of the three baselines. For baseline 5-6 and 6-7 the phases at the zebra stripes (darker colors) and background continuum are evidently different, but there is no notable difference for baseline 7-5.}\label{fig:zb_phz}
\end{center}
\end{figure}

\begin{figure}
\begin{center}
\includegraphics[width=0.85\textwidth]{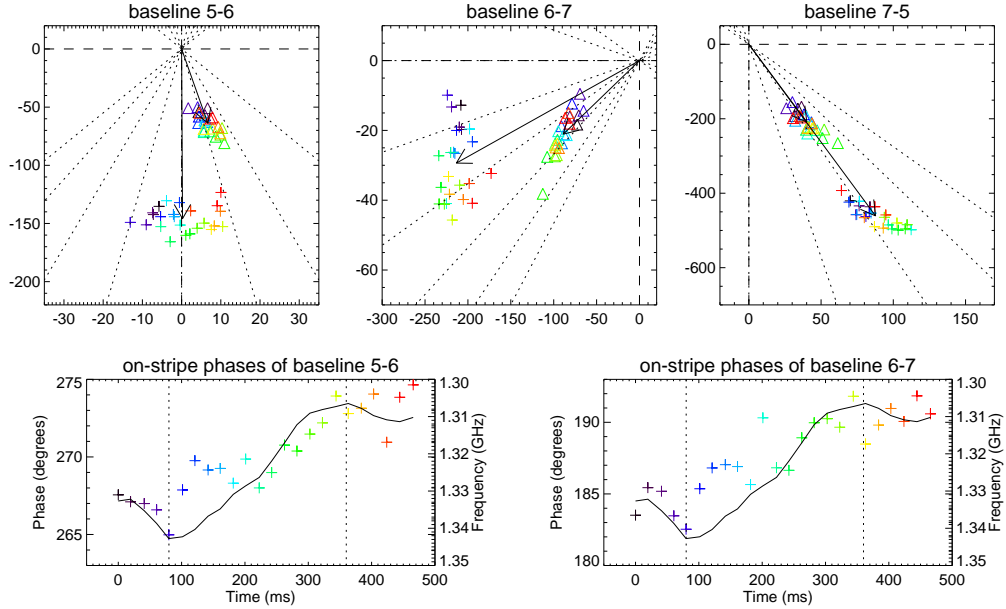}
\caption{The first row: phasor diagrams showing the visibility amplitude and phase variations of ZP for all the three baselines. The visibilities are averaged across the zebra stripes, showing the variations in time (colored from blue to red). The amplitude and phase for each visibility data are represented by its absolute distance from the origin and the direction, respectively. Pluses and triangles denote the on- and off-stripe sources. Arrows give the average amplitudes and phases. The second row: the ZP phases of baselines 5-6 and 6-7 as a function of time. The solid line, repeated in each panel, represents the ZP frequency as a function of time (note that the frequency scale on the right axis is reversed - with frequency decreasing from high to low values). }\label{fig:phasor_t}
\end{center}
\end{figure}

\begin{figure}
\begin{center}
\includegraphics[width=0.85\textwidth]{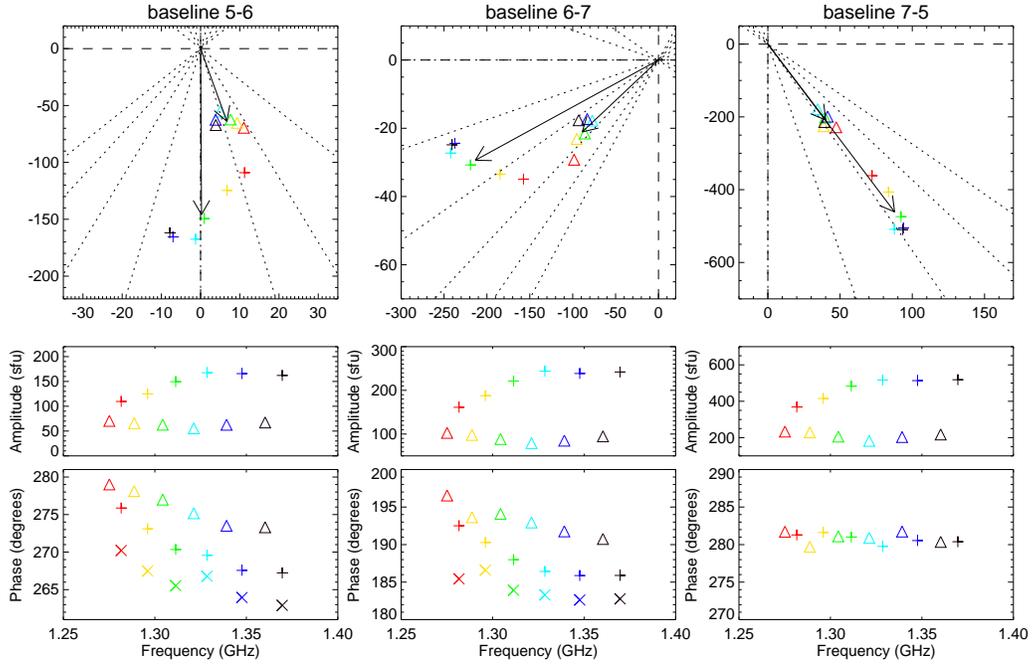}
\caption{The first row: phasor diagrams showing the visibility amplitude and phase variations of ZP for all the three baselines. The visibilities are averaged along the zebra stripes, showing the variations in frequency (colored from black to red for the six stripes with decreasing frequency). Pluses and triangles denote the on- and off-stripe sources. Arrows give the average amplitudes and phases. The second and third rows give the visibility amplitudes and phases of the six stripes. In the first and second panels of the third row, the ``X'' symbols denote the corrected phases of the zebra source. }\label{fig:phasor_f}
\end{center}
\end{figure}

The variation in the on-stripe source sizes seen in Figure \ref{fig:srcsize}b can also be explained by the variation of intensity contrast. The off-stripe source sizes are nearly unaffected by the contrast variation because the continuum source continuously dominates the emission at off-stripe frequencies. But since the on-stripe emission is a mixture of zebra and continuum emission, the variation in contrast plays an important role. At high contrasts (e.g., $P\geqslant 1.5$ at the Nos. 1$-$4 stripes), the zebra source dominates the on-stripe emission, and the source size measurements primarily reflect the property of the zebra source. At lower contrasts (e.g. $P\leqslant 1$ at the Nos. 5$-$6 stripes), the contribution from the continuum source becomes increasingly important and the source size measurements start to reflect the property of the continuum source. Therefore, the on-stripe source sizes start to increase at lower-frequency zebra stripes with decreasing intensity contrasts, and reach the size of the off-stripe source, as shown in Figure \ref{fig:srcsize}b. If we do the vector subtraction on the on-stripe source as above, the corrected relative intensities of zebra source can be obtained, hence their sizes can be fitted accordingly and plotted in Figure \ref{fig:srcsize}b as ``X'' symbols. An up to $20''$ difference between the zebra and continuum (off-stripe) source sizes can be seen after the correction (note the fitted value of the lowest frequency stripe is unreliable, because its relative intensities have large errors that comes from the subtraction of two sources with comparable intensities). We suggest that the size difference is real and indicates the zebra source is yet more compact ($35-40''$) than the source size deduced in \S3.1.2 ($45-50''$) and the lower limit to the ZP brightness temperature should therefore be increased by $\approx 60\%$ to $1.6 \times 10^9$ K.

\subsubsection{Absolute ZP Source Location}

Once calibration of the FST was achieved, as described in \S\ref{sect:instr}, we could locate the absolute source positions by using the three interferometric fringes corresponding to the three antenna baselines to triangulate. The averaged location (in time and frequency) of the ZP source from 22:40:06.86 to 22:40:07.34 UT on the solar disk is shown in Figure \ref{fig:SOTSP} as the intersection of the three fringes, with a dashed circle representing the apparent source size of $\approx\!50''$. The error of the source location is between 1.3$''$ and 3.5$''$ depending on direction. The ZP source is located $\approx\!100''$ to the west of the main flaring emission. For comparison, the OVSA 4.6$-$6.2 GHz map at the same time is over-plotted as contours (the levels have an increment of 10\% of the maximum), showing that the higher-frequency emission is well-correlated with the three Ca II H bright regions (Figure \ref{fig:SOTSP}b). The 4.6$-$6.2 GHz source is far less intense than the ZP emission (cf. Figure \ref{fig:goes_rstn}) and is due to incoherent, non-thermal gyrosynchrotron emission. 

\begin{figure}
\begin{center}
	\includegraphics[width=0.85\textwidth]{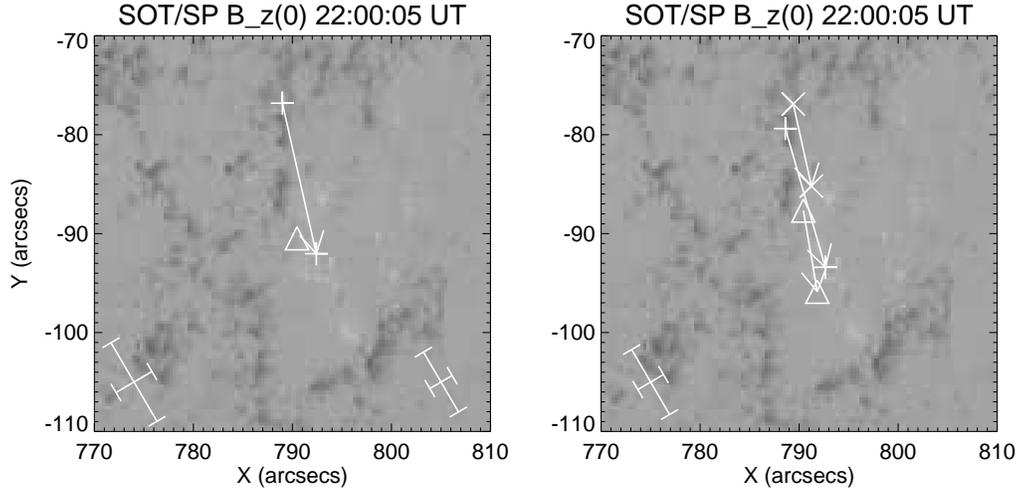}
\caption{(a): The variation of source centroid location along the zebra stripes (with time). The pluses denote the on-stripe source locations along the zebra stripes from 22:40:06.86 to 22:40:07.34 UT (marked in the dashed box in Figure \ref{fig:zebra}), and the arrow shows the position drift. The error is represented by the error bar in the lower-left corner. The triangle denote the off-stripe (continuum) source location, which shows no evident drift, with the error bar plotted in the lower-right corner. (b): the variation of source centroid location across the six zebra stripes (with frequency).The pluses, triangles, and the ``X'' symbol denote respectively the on-stripe, off-stripe and corrected zebra source locations from stripe Nos. 1 to 6 in Figure \ref{fig:zebra} (decreasing in frequency). The arrows give their corresponding position displacements. The error bar in the lower-left corner gives the error.
}\label{fig:zbpos}
\end{center}
\end{figure}

The variation of relative source centroid locations in time and frequency, as was discussed in \S3.1.3, can be now seen from the change of absolute locations on the solar disk. The variation of source location along the zebra stripes -- that is, from 22:40:06.86 to 22:40:07.34 UT --  is shown in Figure \ref{fig:zbpos}a. The pluses denote the on-stripe source locations, with the connecting arrow representing the projected spatial drift of $15.6''$ from NE to SW in 0.48 s. The error bar in the lower-left corner gives the estimated error of the on-stripe location. The off-stripe continuum source location is denoted by a single triangle because it does not display a significant drift in time, with the error bar plotted in the lower-right corner. Figure \ref{fig:zbpos}b shows the source locations as a function of frequency across six zebra stripes. The on- and off-stripe source locations from stripes numbered 1 to 6 (decreasing in frequency) in Figure \ref{fig:zebra} are marked by the pluses and triangles. The ``X'' symbols show the ``actual'' locations of the zebra source after removing the effect of relative intensity variations. The error is denoted by the error bar in the lower-left corner. The connecting arrows show the variation of the source locations with decreasing frequency. We can see that as the frequency decreases, the zebra and continuum (off-stripe) source locations both shift from NE to SW by about $8-9''$, and they are separated from each other by $\approx 11''$. 

\subsection{Magnetic field configuration}
\label{sec:nlfff}

We are interested in constraining the location of the ZP source within the 3D magnetic field configuration. We performed a nonlinear force-free field (NLFFF) extrapolation based on the photospheric vector magnetogram obtained by the SpectroPolarimeter (SP) of the Solar Optical Telescope on board {\sl Hinode}. The SOT/SP measures Stokes profiles of two magnetically sensitive Fe lines at 630.15 and 630.25 nm. We started from the SOT/SP level 2 data that are inverted using the ``MERLIN" inversion code from the polarization spectra. The 180$^{\circ}$ azimuthal ambiguity in the transverse magnetogram was resolved using the ``minimum energy'' method \citep{2006SoPh..237..267M}. The observed vector magnetogram in the image plane (in heliocentric coordinates) was then transformed to heliographic Cartesian coordinates. Because the photospheric magnetic field is not necessarily force-free, an assumption inherent to NLFFF extrapolation as a boundary condition, and since the measurements contain inconsistencies and noise, the measured photospheric magnetic field was preprocessed to mitigate these effects. Here we used the preprocessing method developed by \citet{2006SoPh..233..215W}. Finally, we performed the NLFFF extrapolation using the resulting magnetogram in the heliographic Cartesian coordinates using the weighted optimization method \citep{2004SoPh..219...87W}, which is an implementation of the original method of \citet{2000ApJ...540.1150W}. Best results for a NLFFF extrapolation are achieved when positive and negative magnetic flux are balanced; the FOV we selected to perform the extrapolation is therefore $244''\times 163''$, centered on the active region.  The extrapolation was then calculated on a 240$\times$160$\times$160 grid with a resolution of $1.02''$ in the x-, y- and z-directions (corresponding to the directions of west, north and normal to the tangent plane centered on the active region).

\begin{figure}
\begin{center}
\includegraphics[width=0.9\textwidth]{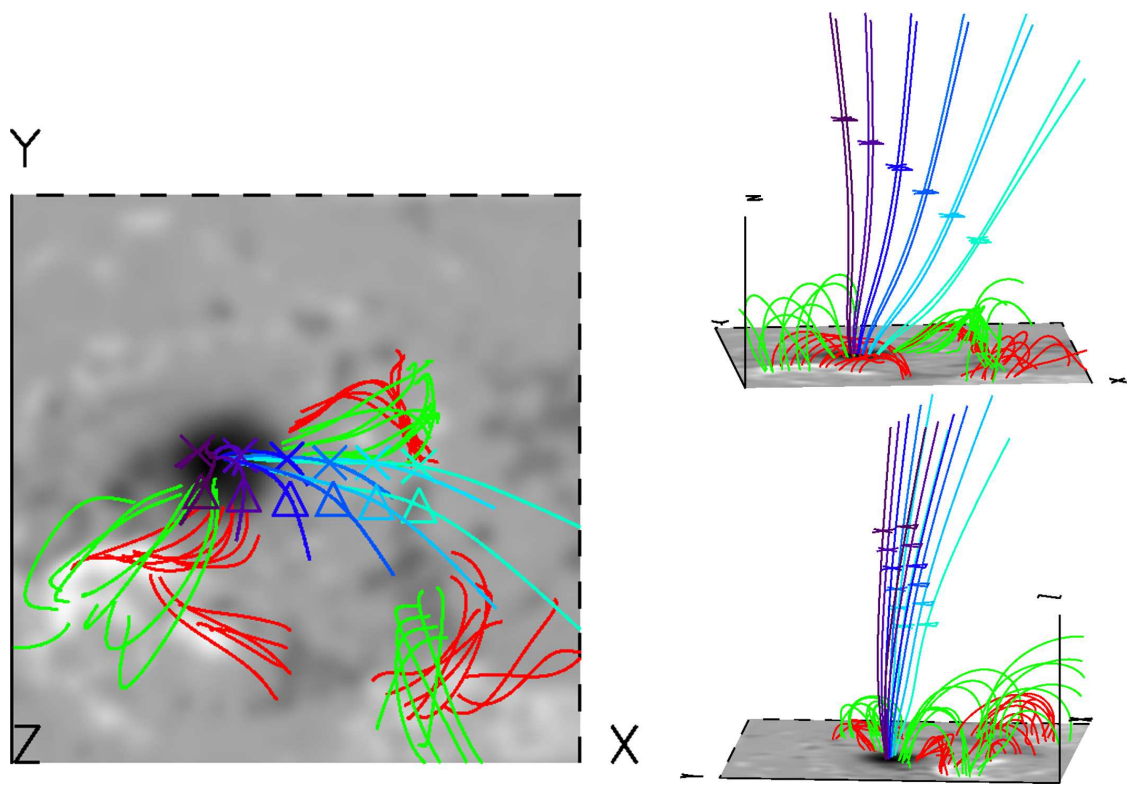}
\caption{The deprojected zebra and continuum source locations with the NLFFF extrapolation field lines. The left image is the view from top, while the two images on the right are for side views from south and east (the x- and y-axis points at west and north, and the z-axis is perpendicular to the tangent plane centered at the active region). The ``X'' symbols and triangles denote a series of possible 3D locations of continuum and zebra sources consistent with their projected locations (averaged in time and frequency) on the image plane. The extrapolated field lines in blue colors passing the possible radio source locations are probably the post-flare loops in which the radio source is located. The colors from light to dark blue denote the magnetic field strengths from 90 to 30 Gauss at heights from 40 to 80 Mm. The magnetic configurations of the three major Ca II H bright regions (see Figure \ref{fig:SOTSP}b) are also shown by the extrapolated field lines, which are grouped and colored in red and green to represent increasing coronal heights.}\label{fig:nlfff}
\end{center}
\end{figure}

In Figure \ref{fig:SOTSP}, we have already seen that the location of the ZP source is nearly $100''$ away from the active region center in the image plane. Since the active region is at S06W46, the ZP source is apparently located relatively high in the solar corona above the active region. The projected ZP source location is known but the location along the line of sight is unknown. We obtained a series of possible 3D locations of the radio source, consistent with its projected location, and plot them in Figure \ref{fig:nlfff}. The zebra and continuum source locations (averaged in time and frequency) are marked as ``X'' symbols and triangles respectively, and color-coded in magnetic field strength (from light to dark blue, the magnetic field changes from $\approx\!90$ G to $\approx\!30$ G at coronal heights from $\approx\!40$ Mm to $\approx\!80$ Mm). It can be seen that as the radio source locations are placed higher in the corona, the magnetic field strength decreases, and they move to a position more nearly over the large sunspot with negative polarity. A group of extrapolated field lines in blue colors is drawn passing these possible source locations, showing a post-flare loop system that is connected with the large sunspot with negative polarity. The polarity of these field lines suggest that since the ZP is RCP, it is polarized in the sense of the ordinary mode. We note, too, that a Ca II H brightening pointed out at the beginning of \S3 (in Figure \ref{fig:ca2h_n_xrt}c) is near the footpoints of the post-flare loops in which the ZP source may be located. These field lines have orientations from NE to SW, and extend to large coronal heights. For completeness, we also show the extrapolated field lines of the three major Ca II H bright regions (see Figure \ref{fig:SOTSP}b), which are grouped and colored in red and green to represent increasing coronal heights. 

\section{Discussion}
\label{sec:discussion}

The key findings of the previous section may be summarized as follows:

\begin{itemize}

\item{An intense X1.5 flare was observed on 2006 Dec 14. The flare was accompanied by intense, prolonged, and variable decimeter wavelength emission as well as gyrosynchrotron emission at centimeter wavelengths. A striking ZP radio burst was observed during the decay phase, centered near 1.32 GHz, with an overall frequency bandwidth of up to 150 MHz. It is completely right-circularly polarized and consists of up to 12 zebra stripes superposed on broadband continuum emission. The stripes drift in the spectrum irregularly as an entity with an average drift rate of $-$50 MHz~s$^{-1}$. The relative frequency separations $\Delta f_s/f_m$ between adjacent stripes are $\approx\!1.3\%$, with $\Delta f_s$ increasing with frequency. }

\item{Two radio sources with a spatial separation of $\approx\!11''$ - zebra and background continuum - contribute to the ZP emission. The apparent zebra source size is around $35''$, which is systematically smaller than that of the continuum source by as much as $20''$, after the correction for the relative intensity effect. The source size of both the ZP and continuum sources are likely strongly affected by scattering. The lower limit of the brightness temperature of ZP is estimated to be $1.6\times 10^9$ K.}

\item{The zebra source centroid drifts irregularly in time with an average drift of $\approx\!16''$ in 0.48 s from NE to SW (corresponding to a projected average velocity of $2.5\times 10^9$ cm~s$^{-1}$), while the continuum locations show no evident drift with time. }

\item{Both the ZP on-stripe and off-stripe source locations are frequency dependent. After correction of the source centroid positions for relative intensity variations, the centroid position shifts $8-9''$ for both the zebra and continuum sources in the direction from NE to SW across the six stripes from high to low frequency at a fixed time. }

\item{The zebra and continuum sources are possibly located in a post-flare loop system with an orientation from NE to SW, which connects to the large sunspot with negative polarity. The polarity of the magnetic field and the observed sense of circular polarization of the ZP imply that it is o-mode. }

\end{itemize}

Of these findings, perhaps the most significant are that the ZP (on-stripe) and continuum (off-stripe) emissions show an angular separation as well as a difference in source size; and that the ZP and continuum source locations are frequency dependent. In other words, the radiation at different frequencies originates from different spatial locations. We are able to conclude that the ZP and continuum sources are not spatially coincident and that both the ZP and continuum sources are spatially extended.

We should point out that the spatial displacement of the on- and off-stripe emissions suggests that the previous measurements of ZP spatial drift in time based on interferometric observations at a single frequency may be misleading. Since the ZP shows frequency drifts with time as an entity, a record of the flux density with time at a single frequency cuts through different zebra stripes as well as the off-stripe continuum source. Apparent spatial shifts can appear as a result of the switch between different sources with different locations. We note that \citet{2005A&A...431.1037A} reported a microwave zebra-pattern structure near 5.7 GHz that was observed on 2003 January 05 by the SBRS and the SSRT. The apparent source size, measured at a frequency far less influenced by scattering, was measured to be $<10''$. The source positions of two successive ZP stripes coincide spatially in the east-west direction. For these reasons, the authors concluded that the emission is due to the nonlinear coupling of Bernstein waves in an unresolved source. Yet in \citet{2005A&A...431.1037A} the spatial coincidence of the two ZP stripes was based on measurements in one dimension at a fixed frequency at different times.  It is possible that the ZP source location drifting with time can result in the apparent ``coincidence'' of the two stripes. Moreover, even if the two stripes do coincide spatially in the east-west direction, there could be an unknown separation in the north-south direction that is not revealed by the one-dimensional measurements. 

In the case of the 14 December 2006 ZP event reported here, and in contrast to the observations of \citet{2005A&A...431.1037A}, there is clear evidence that the source location is frequency-dependent. Taken together with the theoretical difficulties raised by various authors \citep[e.g.][]{2009CEAB...33..281Z}, we conclude that Bernstein models are unlikely to be relevant to this ZP emission and we turn our attention to whistler and DPR models. 

The observed projected average spatial drift velocity of the ZP source along the zebra stripes ($2.5\times 10^9$ cm~s$^{-1}$ in the NE to SW direction) is about 40 times larger than the local Alfv\'{e}n velocity of $\approx\!700$ km~s$^{-1}$ calculated as $V_A=B/(4\pi n_i m_i)^{1/2}\approx 2\times 10^{11} B n_e^{-1/2}$, where $B$ is the magnetic strength in the radio source, assumed here to be $\sim\!50$ G, and $n_e^{1/2}=f/(e^2/\pi m_e)^{1/2}\approx f/9000$ if one assumes the ZP emission at frequency $f$ is near the local plasma frequency. This velocity is too high for most physical movements of the radio source reacting to the magnetic field variation. But it could be explained in terms of the whistler model by propagation of low-frequency whistler wave packets with group velocities ($v_{gr}$) given by \citep{1976SvA....20..582C}
\begin{equation}
v_{gr}=2c\frac{f_{ce}}{f_{pe}}\sqrt{\frac{f_w}{f_{ce}}(1-\frac{f_w}{f_{ce}})^3}
\end{equation}
in the quasi-longitudinal case, where $c$ is the speed of light and $f_w$ is the whistler wave frequency. For $f_w/f_{ce}\approx 0.25$, the ratio at which the growth of the whistler wave is preferred in the decimetric range \citep{2006SSRv..127..195C}, the group velocities can reach the observed value of $2.5\times 10^9$ cm~s$^{-1}$ for $f_{pe}/f_{ce}\lesssim 10$ \citep[see Fig. 2 of][]{1976SvA....20..582C}. 

Is the whistler model able to explain the spatial drift of the ZP source in frequency and time simultaneously? In the whistler model, periodically generated whistler wave packets propagate along a trajectory in the magnetic trap at the whistler group velocity $v_{gr}$. They separate regularly from each other in space so that they can emit at discrete frequencies, thereby producing zebra stripes. The frequency drift of each zebra stripe in the spectrum is determined by the motion of the corresponding whistler wave packet along its trajectory, the density gradient along its trajectory, and wave-wave interactions. We suppose the observed spatial drift with time corresponds to $v_{gr}$. If the coronal number density decreases exponentially with height
\begin{equation}\label{eq:n_prof}
 n_e=n_{e0}e^{-\Delta h/L_n},
\end{equation}
where $n_{e0}$ is the plasma density at the reference height $h_0$, $\Delta h$ is the height from $h_0$ and $L_n=dh/(-dn_e/n_e)$ is the plasma density scale height, then the plasma frequency $f_{pe}$ can be written as
\begin{equation}\label{eq:f_prof}
 f_{pe}=f_{pe0}e^{-\Delta h/2L_n}.
\end{equation}
The radiation is near the plasma frequency, so the frequency drift rate is
\begin{equation}
\dfrac{df}{dt}=-f\dfrac{v_h}{2L_n},
\end{equation}
where $v_h$ is the vertical component of the whistler group velocity (normal to the solar surface). Given the observed average frequency drift rate $df/dt\approx -50$ MHz s$^{-1}$ and the projected velocity of $v_{proj}=2.5\times 10^9$ cm~s$^{-1}$ (which, given its nearly N-S orientation, is nearly parallel to the solar surface), the tangent angle of the trajectory is then $\tan\alpha_1=v_h/v_{proj}=3\times 10^{-11}L_n$. On the other hand, we observe the projected spatial displacement from zebra stripes Nos. 1$-$6 to be $\Delta l_{1-6}=8.5''$, or $6.2\times 10^8$ cm, and their frequencies differ by $\Delta f_{1-6}\approx\!85$ MHz. It's easy to see from Eq. \ref{eq:f_prof} that the height difference between zebra stripes Nos. 1$-$6 $\Delta h_{1-6}\approx\Delta f_{1-6}/f\cdot 2L_n$ for small variations of $f$ and $L_n$. Therefore, we have another estimate of the tangent angle of the whistler wave packet trajectory $\tan\alpha_2=\Delta h_{1-6}/\Delta l_{1-6}\approx 2\times 10^{-10}L_n$, which is an order of magnitude larger than $\tan\alpha_1$ that we obtained based on the assumption that the observed source drift with time corresponded to the whistler group velocity $v_{gr}$. In other words, the trajectory of whistler wave packets cannot be reconciled with both the observed spatial displacement of the ZP source with frequency and the observed frequency drift with time. 

\begin{figure}
\begin{center}
\includegraphics[width=0.9\textwidth]{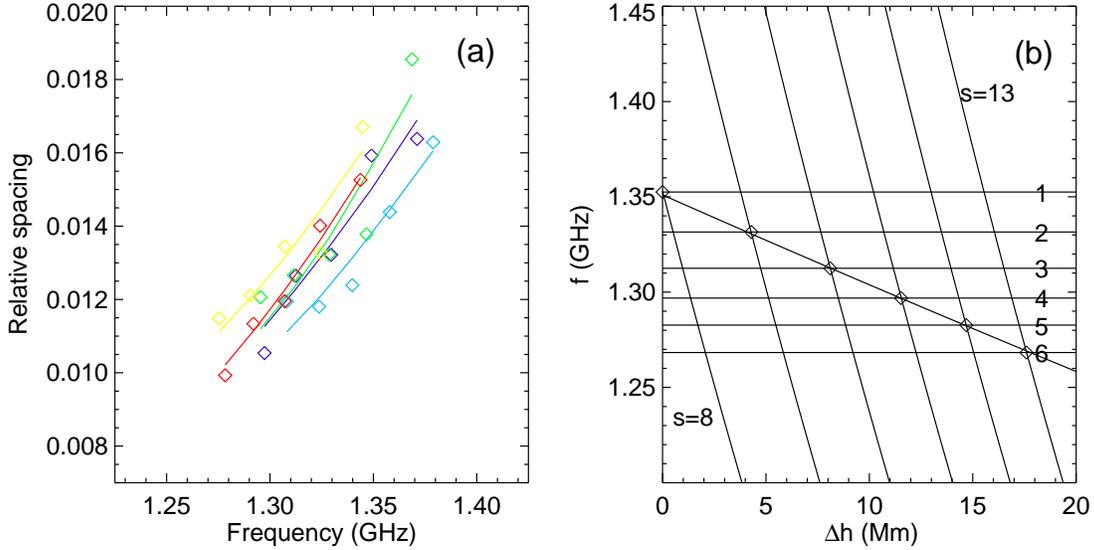}
\caption{(a): Fitting examples of ${\Delta f_s}/{f_m}$ as a function of $f_m$ for the six zebra stripes in the dashed box of Figure \ref{fig:zebra}a (which has 24 consecutive measurements in time from 22:40:06.86 UT to 22:40:07.34 UT). The square symbols and solid lines represent respectively the measured and best-fit values at five near-equally spaced times spanning the entire fitting time range (22:40:06.86 UT, 22:40:06.96 UT, 22:40:07.04 UT, 22:40:07.16 UT, and 22:40:07.26 UT). They are color-coded in time from blue to red. The most probable fitting values of the hamonic numbers are $s=8, 9, 10, 11, 12, 13$ for the six zebra stripes numbered from 1 to 6 in Figure \ref{fig:zebra}a, and the corresponding $L_n/L_B=4.4$. (b) Horizontal lines are the averaged peak frequencies of the six zebra stripes in the solid box of Figure \ref{fig:zebra}a (around 22:40:07.21 UT). The curves of gyroharmonics $sf_{ce}$ with $s=8-13$ are plotted as a function of coronal height $\Delta h$ relative to the height of the $s=8$ layer. The intersections of the curves $sf_{ce}(\Delta h)$ and the plasma frequency distribution $f_{pe}(\Delta h)$ are the DPR levels.}\label{fig:zb_spacing}
\end{center}
\end{figure}

It is also worth noting that the whistler model does not account for the systematic increase of the relative spacing ${\Delta f_s}/{f_m}$ with increasing ZP frequencies as seen in Figure \ref{fig:zb_spacing}a, because the spatial separation between adjacent whistler wave packets is assumed to be result from the periodic injection of whistler wave packets and their frequency separation is therefore essentially constant.

We now consider whether the DPR model is consistent with the observed features of the ZP and continuum source, including both the total power spectral features and their apparent shift in spatial location as a function of time and frequency. As the ZP and continuum sources emit near the local upper hybrid frequency $f_{UH}$ which, with $f_{ce}\ll f_{pe}$, is near the local electron plasma frequency $f_{pe}$,  lower frequency emission should come from greater coronal heights for both the ZP and continuum sources. As a result, a spatial shift with frequency is expected for both the ZP and continuum sources, in accordance with the observations. Furthermore, The NE-SW direction of the spatial drift is generally consistent with the NE-SW orientation of the post-flare loop system in which the emission sources are located. However, the absolute height of the emission source is not yet known. The ZP stripes are those locations where $f_{UH}\approx f_{pe}=sf_{ce}$ in the DPR model. We again assume an exponential dependence of coronal number density as in Eq. \ref{eq:n_prof}, with a scale height $L_n$. The coronal magnetic field is likewise assumed to decrease exponentially with height:
\begin{equation}
 B=B_0e^{-\Delta h/L_B},
\end{equation}
where $B_0$ is the magnetic field at $h_0$ and $L_B$ is the magnetic field scale height. The electron cyclotron frequency $f_{ce}\propto B$ and so
\begin{equation}
 f_{ce}=f_{ce0}e^{-\Delta h/L_B}.
\end{equation}
It can be shown that the relative frequency spacing between the adjacent zebra stripes in the DPR model is then given by \citep{2007SoPh..241..127K}
\begin{equation}\label{eq:DPR}
 \left|\frac{\Delta f_s}{f_m}\right|\simeq \frac{1}{s}\frac{1}{1-(2L_n/L_B)},
\end{equation}
where $s$ is the harmonic number with $f=sf_{ce}$. 
Given the continuity of each zebra stripe in the dynamic spectrum for the $\approx 0.5$ s duration of the spectral fragment in our analysis (the dashed box in Figure \ref{fig:zebra}a), we assume that each stripe corresponds to a single integer harmonic $s$ of the electron cyclotron frequency $f_{ce}$. Moreover, successive zebra stripes emitting at discrete frequencies should have a one-to-one correspondence with successive integer values of $s$, i.e. $s_0$, $s_0+1$, ..., where $s_0$ is the reference harmonic number at $h_0$. Thus the frequency spacing $\Delta f_s/f_m$ is a function of ($s_0+i$), i=0, 1, 2 ... .

By using the observed values of $\Delta f_s/f_m$ for each ZP frequency at a fixed time, we obtain a pair of best-fit values of $s_0$ and $L_n/L_B$. In particular, for a given time, we start with a fixed $s_0$ and perform a least-squares fit to find $L_n/L_B$. Then we increment the value of $s_0$ by integer values until a minimum in the standard deviation of the fit to $\Delta f_s/f_m$ is reached.  We did such fits of $\Delta f_s/f_m$ as a function of ($s_0+i$) for the six zebra stripes for the 24 consecutive integrations shown by the dashed box in Figure \ref{fig:zebra}a. Figure \ref{fig:zb_spacing}a shows examples of these fits at five near-equally spaced times spanning the entire fitting time range (22:40:06.86 UT, 22:40:06.96 UT, 22:40:07.04 UT, 22:40:07.16 UT, and 22:40:07.26 UT), where square symbols and solid lines represent respectively the measured and best-fit values, and are color-coded in time from blue to red. From the distribution of best-fit values of $s_0$, after excluding a few outliers, we conclude that $s_0=8$ is the most probable value, with a scatter of $\sigma_{s_0}\approx\!1.4$. Therefore we assign harmonic numbers of $s=8, 9, 10, 11, 12, 13$ to the six successive zebra stripes numbered from 1 to 6 in Figure \ref{fig:zebra}a, with a corresponding $L_n/L_B\approx\!4.4\pm 0.5$. Note that a stripe with lower frequency corresponds to a higher harmonic number, which is consistent with the previous results based on DPR models \citep[e.g.][]{2003A&A...410.1011Z,2007SoPh..241..127K}. Such a ratio of the scale heights means the magnetic field changes faster than the plasma density with height, which is usually the case in the solar corona. The magnetic field strengths $B$ from stripes 1 to 6 (from low to high in height) can be estimated to be from 62 G to 35 G, by using the harmonic values $s$ and assuming $f\simeq f_{pe}\simeq sf_{ce}$. The uncertainty in $B$ can be estimated from the scatter in the distribution of $s_0$. We find that it ranges from 17\% at $s_0=8$ to 11\% for $s=s_0+5=13$. Using the magnetic fields derived from the NLFFF extrapolation results in \S\ref{sec:nlfff} for guidance, we suggest the zebra stripes 1 to 6 of the ZP source are consistent with a location in the the post-flare loop system at coronal heights from 57 to 75 Mm above the photosphere. The magnetic scale height $L_B$ can be estimated to be $L_B=dh/(-dB/B)\approx 3.2\times 10^9$ cm. On the other hand, an estimation of the density scale height $L_n$ is also available by $L_n=dh/(-2df/f)\approx 1.4\times 10^{10}$ cm, from the known frequencies of the zebra stripes (an equivalent result of $L_n$ can be obtained by using $L_B$ and the fitted value of $L_n/L_B\approx 4.4$). 

DPR levels are formed as a result of plasma density and magnetic field variations in height, as demonstrated by Figure \ref{fig:zb_spacing}b. By applying the values of $L_n$, $L_B$, and $s$, the dependencies of $sf_{ce}$ and $f_{pe}$ as a function of $\Delta h$ are given for the time denoted as the solid box in Figure \ref{fig:zebra}a. The reference height $h_0$ is set to be that of the lowest DPR level (the No. 1 stripe with $s_0=8$). The intersections of the curves $sf_{ce}(\Delta h)$ and the plasma frequency distribution $f_{pe}(\Delta h)$ (marked by diamonds) are the DPR levels, which coincide with the observed frequencies of zebra stripes (denoted by the horizontal lines). In this event, the zebra stripes drift irregularly and rapidly in the dynamic spectrum, indicating that the DPR levels do not remain at precisely the same height in the flare loop - that is, $h_0$ varies by a few percent of $L_n$, or a few thousand kilometers - which also accounts for the scatter and temporal variation seen in Figure \ref{fig:zb_spacing}a. 

In the whistler model, formation of the regularly spaced stripes of ZP in frequency is based on propagation of the periodically generated whistler wave packets along a given trajectory - that is, the trajectory of each whistler wave packet determines the spatial drift of each zebra stripe in time (with tangent angle $\tan\alpha_1$), should follow the same orientation of the spatial extent of the entire ZP source, represented by the spatial displacement of zebra stripes at different frequencies (with  tangent angle $\tan\alpha_2$), i.e. $\tan\alpha_1 \approx \tan\alpha_2$. We have already demonstrated that they can not be reconciled in the context of the whistler model. For the DPR model the two ``trajectories'' need not coincide.  The DPR levels, corresponding to the observed zebra stripes, are distributed along a coronal loop at locations where the resonant conditions are matched. At the same time, the locations of each DPR level can change with time in response to the variation of source conditions. Therefore, the ``trajectory'' of each DPR level is not required to follow the same orientation of the spatial distribution of the resonance layes in the loop. In fact, given the density scale height of $L_n\approx 1.4\times 10^{10}$ cm, we have $\alpha_1\approx 23^{\circ}$ and $\alpha_2\approx 70^{\circ}$. That means the DPR levels are distributed near vertically in the coronal loop while their apparent motion is nearly horizontal in time. 

Both the ZP and continuum sources probably reside on the same post-flare/post-CME loops that extend field lines from in or near the large sunspot with negative polarity to well up into the corona with a NE to SW orientation. The radio emission could be powered by an energy release site high up in the corona above the radio source. This site could be related to magnetic reconnections induced by the fast halo CME associated with this flare, which may also account for the intense and prolonged type IV burst activity in the post-flare phase, from 22:07-23:15 UT. As mentioned in \S\ref{sec:nlfff}, the radio source may be magnetically associated with the Ca II H brightening feature that persists for the duration of the type IV emission. The brightening may indicate the magnetic footpoints of the field lines on which the ZP and type IV emission originates. 

In the DPR model, ZPs are thought to arise from energetic electrons in a magnetic trap with a large gradient in electron momentum distribution function perpendicular to the magnetic field $\partial f/\partial p_\bot$, as might arise from a distribution function sharply peaked perpendicular to the magnetic field (e.g., a Dory-Guest-Harris, or ring,  distribution), or a loss-cone distribution with sufficiently narrow momentum dispersion \citep[e.g.][]{1986ApJ...307..808W,2007SoPh..241..127K,2009SoPh..255..273Z}. At the same time, the continuum emission can also arise from an anisotropic electron distribution in the magnetic trap, most likely, a common loss-cone distribution \citep[e.g.][]{2007SoPh..241..127K,2009SoPh..255..273Z}. Therefore, it is plausible that the ZP and continuum observed by the FST is related to the injection of fast electrons that originate from the energy release site above the radio source, perhaps as the result of magnetic reconnection behind the fast halo CME associated with the flare. These downward-propagating electrons establish the anisotropic distribution that drives the zebra and continuum emission. It is highly unlikely that the ambient electron number density or the magnetic field change significantly during the $\sim 1$ s duration of the ZP event reported here and therefore cannot be the reason for variations in the ZP frequency drift with time. More likely, variations in the number and/or degree of anisotropy of the injected electrons lead to variations in the height where wave growth is favored and the DPR condition is met. As a result, the heights of the established anisotropic distribution and/or the DPR levels can be modulated, and form the irregular ZP frequency drift in the dynamic spectrum. The observed irregular drift of ZP source centroid in time can also be attributed to this effect.

As has been proposed by many authors \citep[e.g.][]{1986ApJ...307..808W,2007SoPh..241..127K,2009SoPh..255..273Z}, the conditions required to produce a ZP signature through the DPR instability are more stringent than those for producing a continuum, e.g., a low-momentum electron deficit, a narrower momentum dispersion, a higher overall momentum, etc. In addition, for achieving the observed large intensity contrast of the ZP to the continuum in this event, the growth rate of the ZP emission should be sufficiently higher than that of the continuum, which may require a higher non-equilibrium electron density. All the peculiarities mentioned above could possibly explain the observations of the relatively short-lived and fast-varying ZP and the more persistent and stable continuum in this event. 

\section{Conclusion}

We present FST observations of a striking zebra pattern radio burst that occurred during the 14 December 2006 X3.4 flare. This is the first observation of zebra pattern emission that combines simultaneous high spectral and time resolution data with interferometric observations over the entire bandwidth. After calibrating the FST against OVSA we can obtain the absolute locations of radio fine structures on the solar disk, and study their spatial and spectral features. We conclude that the DPR model is the most favorable model since it can fit most spectral and spatial features of this ZP event. 

\begin{figure}
\begin{center}
\includegraphics[width=0.8\textwidth]{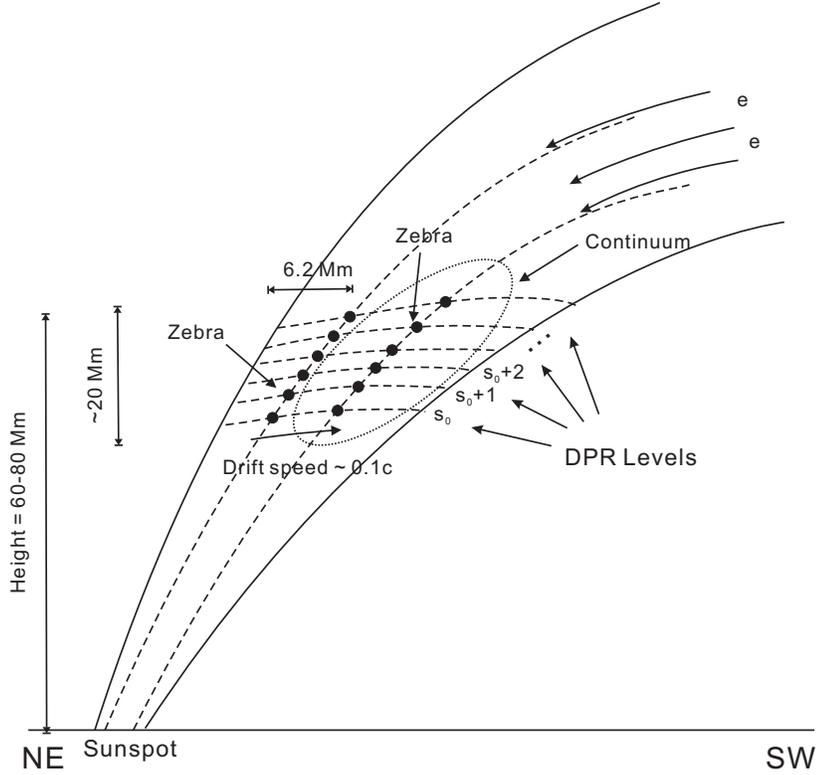}
\caption{A simplified source model: the zebra and continuum source are located at a height of $60-80$ Mm in a post-flare/post-CME loop system that connects the large sunspot with negative polarity with a NE to SW orientation. Both the zebra and continuum are extended sources occupying a total height range of $\approx\!20$ Mm in the post-flare loops. Within the zebra source, individual stripes correspond to emissions near the local plasma frequencies at the DPR levels (horizontal dashed lines, of which the lowest one corresponds to the $s_0 = f_{pe}/f_{ce} = 8$ layer). An energy release site is located high up in the corona above the radio source. Electron beams generated from this site propagate downwards along the magnetic field lines into the magnetic trap and give rise to the instability for emission. The fast and irregular spatial drift of the ZP source centroid in time ($\approx 0.1c$, indicated by the arrow showing the general drift direction of NE$-$SW) and the irregular frequency drift of ZP likely result from time variations in properties of the fast electrons injected into the field well above the ZP source. The continuum source is comparatively more extended in size, with its emission centroid separated from that of the zebra source in the NE$-$SW direction.}\label{fig:cartoon}
\end{center}
\end{figure}

The cartoon in Figure \ref{fig:cartoon} summarizes our interpretation of the ZP event within the framework of DPR model: the zebra and continuum source are located at a height of $60-80$ Mm in a post-flare/post-CME loop system that connects the large sunspot with negative polarity with a NE to SW orientation. Both the zebra and continuum are extended sources occupying a total height range of $\approx\!20$ Mm in the post-flare loops, which explains the spatial drifts of ZP and continuum in frequency. Within the zebra source, individual stripes correspond to emissions near the local plasma frequencies at the DPR levels. We suggest that the fast electrons responsible for the continuum and ZP emission originate in an energy release site high up in the corona above the radio source, perhaps the result of magnetic reconnections induced by the fast halo CME. The fast and irregular spatial drift of the ZP source centroid in time and the irregular frequency drift of ZP likely result from time variations in properties of the fast electrons injected into the field well above the ZP source. The continuum source is comparatively more extended in size, with its emission centroid separated from that of the zebra source in the NE$-$SW direction. The continuum emission requires less stringent conditions for the anisotropic distribution of the injected electrons and can therefore have a different size and a different source centroid location. This may also explain why the continuum emission is comparatively more stable in time and frequency.

\acknowledgments

\noindent The National Radio Astronomy Observatory is a facility of the National Science Foundation operated under cooperative agreement by Associated Universities, Inc. {\it Hinode} is a Japanese mission developed and launched by ISAS/JAXA, with NAOJ as a domestic partner and NASA and STFC (UK) as international partners. It is operated by these agencies in co-operation with ESA and NSC (Norway). OVSA and FST are supported by NSF grant AST-0908344 to New Jersey Institute of Technology. Bin Chen is supported by NSF grant AGS-1010652 and Ju Jing is supported by NSF grant ATM 09-36665.

\bibliographystyle{apj}
\bibliography{apj-jour,FSTpaper}

\begin{thebibliography}{46}
\expandafter\ifx\csname natexlab\endcsname\relax\def\natexlab#1{#1}\fi

\bibitem[{{Altyntsev} {et~al.}(2005){Altyntsev}, {Kuznetsov}, {Meshalkina},
  {Rudenko}, \& {Yan}}]{2005A&A...431.1037A}
{Altyntsev}, A.~T., {Kuznetsov}, A.~A., {Meshalkina}, N.~S., {Rudenko}, G.~V.,
  \& {Yan}, Y. 2005, \aap, 431, 1037

\bibitem[{{Aurass} {et~al.}(2003){Aurass}, {Klein}, {Zlotnik}, \&
  {Zaitsev}}]{2003A&A...410.1001A}
{Aurass}, H., {Klein}, K., {Zlotnik}, E.~Y., \& {Zaitsev}, V.~V. 2003, \aap,
  410, 1001

\bibitem[{{B{\'a}rta} \& {Karlick{\'y}}(2006)}]{2006A&A...450..359B}
{B{\'a}rta}, M., \& {Karlick{\'y}}, M. 2006, \aap, 450, 359

\bibitem[{{Bastian}(1994)}]{1994ApJ...426..774B}
{Bastian}, T.~S. 1994, \apj, 426, 774

\bibitem[{{Bastian}(2004)}]{2004ASSL..314.....G}
---. 2004, {Solar and Space Weather Radiophysics - Current Status and Future
  Developments}, ed. {D.~E.~Gary \& C.~U.~Keller} ({Boston: Kluwer Academic
  Publishers}), 47--69

\bibitem[{{Chen} \& {Yan}(2007)}]{2007SoPh..246..431C}
{Chen}, B., \& {Yan}, Y. 2007, \solphys, 246, 431

\bibitem[{{Chernov}(1976)}]{1976SvA....20..582C}
{Chernov}, G.~P. 1976, \sovast, 20, 582

\bibitem[{{Chernov}(1990)}]{1990SoPh..130...75C}
---. 1990, \solphys, 130, 75

\bibitem[{{Chernov}(2005)}]{2005PlPhR..31..314C}
---. 2005, Plasma Physics Reports, 31, 314

\bibitem[{{Chernov}(2006)}]{2006SSRv..127..195C}
---. 2006, \ssr, 127, 195

\bibitem[{{Chernov} {et~al.}(1994){Chernov}, {Klein}, {Zlobec}, \&
  {Aurass}}]{1994SoPh..155..373C}
{Chernov}, G.~P., {Klein}, K., {Zlobec}, P., \& {Aurass}, H. 1994, \solphys,
  155, 373

\bibitem[{{Chernov} {et~al.}(1998){Chernov}, {Markeev}, {Poquerusse},
  {Bougeret}, {Klein}, {Mann}, {Aurass}, \& {Aschwanden}}]{1998A&A...334..314C}
{Chernov}, G.~P., {Markeev}, A.~K., {Poquerusse}, M., {Bougeret}, J.~L.,
  {Klein}, K., {Mann}, G., {Aurass}, H., \& {Aschwanden}, M.~J. 1998, \aap,
  334, 314

\bibitem[{{Chernov} {et~al.}(2001){Chernov}, {Poquerusse}, {Bougeret}, \&
  {Zlobec}}]{2001RaSc...36.1745C}
{Chernov}, G.~P., {Poquerusse}, M., {Bougeret}, J., \& {Zlobec}, P. 2001, Radio
  Science, 36, 1745

\bibitem[{{Chernov} {et~al.}(2006){Chernov}, {Sych}, {Yan}, {Fu}, {Tan},
  {Huang}, {Wang}, \& {Wu}}]{2006SoPh..237..397C}
{Chernov}, G.~P., {Sych}, R.~A., {Yan}, Y., {Fu}, Q., {Tan}, C., {Huang}, G.,
  {Wang}, D., \& {Wu}, H. 2006, \solphys, 237, 397

\bibitem[{{Chernov} {et~al.}(2003){Chernov}, {Yan}, \&
  {Fu}}]{2003A&A...406.1071C}
{Chernov}, G.~P., {Yan}, Y.~H., \& {Fu}, Q.~J. 2003, \aap, 406, 1071

\bibitem[{{Chernov} {et~al.}(2005){Chernov}, {Yan}, {Fu}, \&
  {Tan}}]{2005A&A...437.1047C}
{Chernov}, G.~P., {Yan}, Y.~H., {Fu}, Q.~J., \& {Tan}, C.~M. 2005, \aap, 437,
  1047

\bibitem[{{Elgar{\"o}y}(1959)}]{1959Natur.184..887E}
{Elgar{\"o}y}, O. 1959, \nat, 184, 887

\bibitem[{{Elgar{\o}y}(1982)}]{1982ITABO..53.....E}
{Elgar{\o}y}, {\O}. 1982, Inst.~Theor.~Astrophys., Blindern-Oslo, Rep., No.~53,
  30 pp., 53

\bibitem[{{Fleishman} {et~al.}(1994){Fleishman}, {Stepanov}, \&
  {Yurovsky}}]{1994SoPh..153..403F}
{Fleishman}, G.~D., {Stepanov}, A.~V., \& {Yurovsky}, Y.~F. 1994, \solphys,
  153, 403

\bibitem[{{Gary} \& {Hurford}(1994)}]{1994ApJ...420..903G}
{Gary}, D.~E., \& {Hurford}, G.~J. 1994, \apj, 420, 903

\bibitem[{{Golub} {et~al.}(2007){Golub}, {Deluca}, {Austin}, {Bookbinder},
  {Caldwell}, {Cheimets}, {Cirtain}, {Cosmo}, {Reid}, {Sette}, {Weber},
  {Sakao}, {Kano}, {Shibasaki}, {Hara}, {Tsuneta}, {Kumagai}, {Tamura},
  {Shimojo}, {McCracken}, {Carpenter}, {Haight}, {Siler}, {Wright}, {Tucker},
  {Rutledge}, {Barbera}, {Peres}, \& {Varisco}}]{2007SoPh..243...63G}
{Golub}, L., {et~al.} 2007, \solphys, 243, 63

\bibitem[{{Kosugi} {et~al.}(2007){Kosugi}, {Matsuzaki}, {Sakao}, {Shimizu},
  {Sone}, {Tachikawa}, {Hashimoto}, {Minesugi}, {Ohnishi}, {Yamada}, {Tsuneta},
  {Hara}, {Ichimoto}, {Suematsu}, {Shimojo}, {Watanabe}, {Shimada}, {Davis},
  {Hill}, {Owens}, {Title}, {Culhane}, {Harra}, {Doschek}, \&
  {Golub}}]{2007SoPh..243....3K}
{Kosugi}, T., {et~al.} 2007, \solphys, 243, 3

\bibitem[{{Kuijpers}(1975)}]{1975PhDT.........1K}
{Kuijpers}, J.~M.~E. 1975, PhD thesis, Utrecht, Rijksuniversiteit, Doctor in de
  Wiskunde en Natuurwetenschappen Dissertation, 1975.~72 p.~Research supported
  by the Nederlandse Organisatie voor Zuiver-Wetenschappelijk Onderzoek.

\bibitem[{{Kuznetsov}(2008)}]{2008SoPh..253..103K}
{Kuznetsov}, A.~A. 2008, \solphys, 253, 103

\bibitem[{{Kuznetsov} \& {Tsap}(2007)}]{2007SoPh..241..127K}
{Kuznetsov}, A.~A., \& {Tsap}, Y.~T. 2007, \solphys, 241, 127

\bibitem[{{LaBelle} {et~al.}(2003){LaBelle}, {Treumann}, {Yoon}, \&
  {Karlicky}}]{2003ApJ...593.1195L}
{LaBelle}, J., {Treumann}, R.~A., {Yoon}, P.~H., \& {Karlicky}, M. 2003, \apj,
  593, 1195

\bibitem[{{Laptukhov} \& {Chernov}(2006)}]{2006PlPhR..32..866L}
{Laptukhov}, A.~I., \& {Chernov}, G.~P. 2006, Plasma Physics Reports, 32, 866

\bibitem[{{Ledenev} {et~al.}(2006){Ledenev}, {Yan}, \&
  {Fu}}]{2006SoPh..233..129L}
{Ledenev}, V.~G., {Yan}, Y., \& {Fu}, Q. 2006, \solphys, 233, 129

\bibitem[{{Liu}(2007)}]{2007PhDT....Liu}
{Liu}, Z. 2007, PhD thesis, New Jersey Institute of Technology

\bibitem[{{Liu} {et~al.}(2007){Liu}, {Gary}, {Nita}, {White}, \&
  {Hurford}}]{2007PASP..119..303L}
{Liu}, Z., {Gary}, D.~E., {Nita}, G.~M., {White}, S.~M., \& {Hurford}, G.~J.
  2007, \pasp, 119, 303

\bibitem[{{Metcalf} {et~al.}(2006){Metcalf}, {Leka}, {Barnes}, {Lites},
  {Georgoulis}, {Pevtsov}, {Balasubramaniam}, {Gary}, {Jing}, {Li}, {Liu},
  {Wang}, {Abramenko}, {Yurchyshyn}, \& {Moon}}]{2006SoPh..237..267M}
{Metcalf}, T.~R., {et~al.} 2006, \solphys, 237, 267

\bibitem[{{Rosenberg}(1972)}]{1972SoPh...25..188R}
{Rosenberg}, H. 1972, \solphys, 25, 188

\bibitem[{{Slottje}(1972)}]{1972SoPh...25..210S}
{Slottje}, C. 1972, \solphys, 25, 210

\bibitem[{{Su} {et~al.}(2007){Su}, {Golub}, {van Ballegooijen}, {Deluca},
  {Reeves}, {Sakao}, {Kano}, {Narukage}, \& {Shibasaki
  Kiyoto}}]{2007PASJ...59S.785S}
{Su}, Y., {et~al.} 2007, \pasj, 59, 785

\bibitem[{{Tan}(2010)}]{2010Ap&SS.325..251T}
{Tan}, B. 2010, \apss, 325, 251

\bibitem[{{Tsuneta} {et~al.}(2008){Tsuneta}, {Ichimoto}, {Katsukawa}, {Nagata},
  {Otsubo}, {Shimizu}, {Suematsu}, {Nakagiri}, {Noguchi}, {Tarbell}, {Title},
  {Shine}, {Rosenberg}, {Hoffmann}, {Jurcevich}, {Kushner}, {Levay}, {Lites},
  {Elmore}, {Matsushita}, {Kawaguchi}, {Saito}, {Mikami}, {Hill}, \&
  {Owens}}]{2008SoPh..249..167T}
{Tsuneta}, S., {et~al.} 2008, \solphys, 249, 167

\bibitem[{{Wheatland} {et~al.}(2000){Wheatland}, {Sturrock}, \&
  {Roumeliotis}}]{2000ApJ...540.1150W}
{Wheatland}, M.~S., {Sturrock}, P.~A., \& {Roumeliotis}, G. 2000, \apj, 540,
  1150

\bibitem[{{Wiegelmann}(2004)}]{2004SoPh..219...87W}
{Wiegelmann}, T. 2004, \solphys, 219, 87

\bibitem[{{Wiegelmann} {et~al.}(2006){Wiegelmann}, {Inhester}, \&
  {Sakurai}}]{2006SoPh..233..215W}
{Wiegelmann}, T., {Inhester}, B., \& {Sakurai}, T. 2006, \solphys, 233, 215

\bibitem[{{Winglee} \& {Dulk}(1986)}]{1986ApJ...307..808W}
{Winglee}, R.~M., \& {Dulk}, G.~A. 1986, \apj, 307, 808

\bibitem[{{Zaitsev} \& {Stepanov}(1983)}]{1983SoPh...88..297Z}
{Zaitsev}, V.~V., \& {Stepanov}, A.~V. 1983, \solphys, 88, 297

\bibitem[{{Zheleznyakov} \&
  {Zlotnik}(1975{\natexlab{a}})}]{1975SoPh...43..431Z}
{Zheleznyakov}, V.~V., \& {Zlotnik}, E.~I. 1975{\natexlab{a}}, \solphys, 43,
  431

\bibitem[{{Zheleznyakov} \&
  {Zlotnik}(1975{\natexlab{b}})}]{1975SoPh...44..461Z}
{Zheleznyakov}, V.~V., \& {Zlotnik}, E.~Y. 1975{\natexlab{b}}, \solphys, 44,
  461

\bibitem[{{Zlotnik}(2009)}]{2009CEAB...33..281Z}
{Zlotnik}, E.~Y. 2009, Central European Astrophysical Bulletin, 33, 281

\bibitem[{{Zlotnik} {et~al.}(2009){Zlotnik}, {Zaitsev}, {Aurass}, \&
  {Mann}}]{2009SoPh..255..273Z}
{Zlotnik}, E.~Y., {Zaitsev}, V.~V., {Aurass}, H., \& {Mann}, G. 2009, \solphys,
  255, 273

\bibitem[{{Zlotnik} {et~al.}(2003){Zlotnik}, {Zaitsev}, {Aurass}, {Mann}, \&
  {Hofmann}}]{2003A&A...410.1011Z}
{Zlotnik}, E.~Y., {Zaitsev}, V.~V., {Aurass}, H., {Mann}, G., \& {Hofmann}, A.
  2003, \aap, 410, 1011

\end{thebibliography}

\newpage

\end{document}